\documentclass[11pt,english]{article}

\usepackage{fullpage}
 
\usepackage{amsmath}
\usepackage{amsthm}
\usepackage{amssymb}
\usepackage{hyperref}

\usepackage{microtype}

\usepackage[noend]{algpseudocode}
\usepackage{algorithm}
\usepackage{graphicx}

\usepackage{tikz}
\usepackage{calc}

\usepackage{subcaption}

\usepackage{color}

\usepackage{wrapfig}
\usepackage{xspace}

\usepackage{varioref} 

\usepackage{footnote} 

\usepackage{enumitem} 

\mathchardef\mhyphen="2D

\newcommand{\eps}{\ensuremath{\varepsilon}}

\newcommand{\R}{{\mathbb{R}}}

\newcommand{\thmref}[1]{Theorem~\ref{thm:#1}}

\newcommand{\lemref}[1]{Lemma~\ref{lem:#1}}
\newcommand{\lemrefs}[2]{Lemmas~\ref{lem:#1} and~\ref{lem:#2}}

\newcommand{\corref}[1]{Corollary~\ref{cor:#1}}

\newcommand{\secref}[1]{Section~\ref{sec:#1}}

\newcommand{\eq}[1]{equation~\eqref{eq:#1}}

\newtheorem{thm}{Theorem}[section]
\newtheorem{lem}[thm]{Lemma}

\newtheorem{cor}[thm]{Corollary}
\newtheorem{defn}[thm]{Definition}

\newtheorem{claim}[thm]{Claim}

\newtheorem{hypo}[thm]{Hypothesis}

\global\long\def\Oh{{O}}

\newcommand{\poly}{\textup{poly}}

\newcommand{\dfr}{\ensuremath{d_{\textup{F}}}}
\newcommand{\Fr}{Fr\'echet }

\newcommand{\F}{{F}}

\newcommand{\true}{\mathsf{true}}

\newcommand{\SuccPair}{Formula-Pair\xspace} 
\newcommand{\LCS}{\textup{LCS}\xspace}

\newcommand{\depth}{\textup{depth}}

\newcommand{\defproblem}[4]{
\bigskip
\begin{center}
\noindent\fbox{
	\begin{minipage}{.96\linewidth}
	\textsc{#1}
	
	\smallskip
	\noindent\begin{tabular}{@{}l@{ }l}
	\emph{Input:} & \begin{minipage}[t]{\linewidth-\widthof{Parameter:\ \ \ \ }} #2\end{minipage}\\
	\emph{Question:} & \begin{minipage}[t]{\linewidth-\widthof{Parameter:\ \ \ \ }} #3\end{minipage}\\
	\emph{Complexity:} & \begin{minipage}[t]{\linewidth-\widthof{Parameter:\ \ \ \ }} #4\end{minipage}
	\end{tabular}
	\end{minipage}
}
\end{center}
\medskip
}

\title{Tighter Connections Between Formula-SAT and Shaving Logs\footnote{Part of the work was performed while visiting the Simons Institute for the Theory of Computing, Berkeley, CA. The work was completed when A.A. was at Stanford University and was supported by Virginia Vassilevska Williams' NSF Grants CCF-1417238 and CCF-1514339, and BSF Grant BSF:2012338. 
}}

\author{
Amir Abboud\thanks{IBM Almaden Research Center, \texttt{abboud@cs.stanford.edu}} \and
Karl Bringmann\thanks{Max Planck Institute for Informatics, Saarland Informatics Campus, Germany, \texttt{kbringma@mpi-inf.mpg.de}}
}

\date{}

\begin{document}

\maketitle

\medskip

\begin{abstract}

A noticeable fraction of Algorithms papers in the last few decades improve the running time of well-known algorithms for fundamental problems by logarithmic factors.
For example, the $\Oh(n^2)$ dynamic programming solution to the Longest Common Subsequence problem (LCS) was improved to $O(n^2/\log^{2}n)$ in several ways and using a variety of ingenious tricks.
This line of research, also known as \emph{the art of shaving log factors}, lacks a tool for proving negative results.
Specifically, how can we show that it is unlikely that LCS can be solved in time $O(n^2/\log^3n)$?

Perhaps the only approach for such results was suggested in a recent paper of Abboud, Hansen, Vassilevska W. and Williams (STOC'16).
The authors blame the hardness of shaving logs on the hardness of solving satisfiability on boolean formulas (Formula-SAT) faster than exhaustive search.
They show that an $O(n^2/\log^{1000} n)$ algorithm for LCS would imply a major advance in circuit lower bounds. 
Whether this approach can lead to tighter barriers was unclear.

In this paper, we push this approach to its limit and, in particular, prove that a well-known barrier from complexity theory stands in the way for shaving five additional log factors for fundamental combinatorial problems. 
For LCS, regular expression pattern matching, as well as the Fr\'echet distance problem from Computational Geometry, we show that an $O(n^2/\log^{7+\eps}{n})$ runtime would imply new Formula-SAT algorithms.

Our main result is a reduction from SAT on formulas of size $s$ over $n$ variables to LCS on sequences of length $N=2^{n/2} \cdot s^{1+o(1)}$.  Our reduction is essentially as efficient as possible, and it greatly improves the previously known reduction for LCS with $N=2^{n/2} \cdot s^{c}$, for some $c \geq 100$.

\end{abstract}


\section{Introduction}

Since the early days of Algorithms research, a noticeable fraction of papers each year \emph{shave log factors} for fundamental problems: they reduce the best known upper bound on the time complexity from $T(n)$ to $T(n)/\log^c{n}$, for some $c>0$.
While in some cases a cynic would call such results ``hacks'' and ``bit tricks'', there is no doubt that they often involve ingenious algorithmic ideas and suggest fundamental new ways to look at the problem at hand.
In his survey, Timothy Chan calls this kind of research ``The Art of Shaving Logs'' \cite{Chan13}.
In many cases, we witness a race of shaving logs for some problem, in which a new upper bound is found every few months, without giving any hints on when this race is going to halt. 
For example, in the last few years, the upper bound for combinatorial Boolean Matrix Multiplication dropped from $O(n^3/\log^2 n)$ \cite{FourRussians}, to $O(n^3/\log^{2.25} n)$ \cite{BW09}, to $O(n^3/\log^{3} n)$ \cite{Chan15}, and most recently to $O(n^3/\log^{4}n)$ \cite{Yu15}.
Perhaps the single most important missing technology for this kind of research is a tool for proving \emph{lower bounds}.

Consider the problem of computing the Longest Common Subsequence (LCS) of two strings of length $n$.
LCS has a simple $O(n^2)$ time dynamic programming algorithm \cite{WF74,CLRSbook}. 
Several approaches have been utilized in order to shave log factors such as the ``Four Russians'' technique \cite{FourRussians,HPV77,MP80,BF08,Grabo14}, utilizing bit-parallelism \cite{AD86,CIPR01,Hyyro04}, and working with compressed strings \cite{CLZ03,Gaw12}. 
The best known upper bounds are $O(n^2/\log^2{n})$ for constant size alphabets \cite{MP80}, and $O(n^2\log\log{n}/\log^2{n})$ for large alphabets \cite{Grabo14}.
But can we do better? Can we solve LCS in $O(n^2/\log^3 n)$ time?
While the mathematical intrigue is obvious, we remark that even such mild speedups for LCS could be significant in practice.
Besides its use as the \emph{diff} operation in unix, LCS is at the core of highly impactful similarity measures between biological data.
A heuristic algorithm called BLAST for a generalized version of LCS (namely, the Local Alignment problem \cite{SW81}) has been cited more than sixty thousand times \cite{BLAST}. 
While such heurisitics are much faster than the near-quadratic time algorithms above, they are not guaranteed to return an optimal solution and are thus useless in many applications, and biologists often fall back to (highly optimized implementations of) the quadratic solutions,  see, e.g.~\cite{LST07,CUDASW}. 

How would one show that it is hard to shave logs for some problem?
A successful line of work, inspired by NP-hardness, utilizes ``fine-grained reductions'' to prove statements of the form: a small improvement over the known runtime for problem A implies a breakthrough algorithm for problem B, refuting a plausible hypothesis about the complexity of B.
For example, it has been shown that if LCS can be solved in $O(n^{2-\eps})$ time, where $\eps>0$, then there is a breakthrough $(2-\delta)^n$ algorithm for CNF-SAT, and the Strong Exponential Time Hypothesis (SETH, defined below) is refuted \cite{ABV15,BK15}.
Another conjecture that has been used to derive interesting lower bounds states that the $3$-SUM problem\footnote{$3$-SUM asks, given a list of $n$ numbers, to find three that sum to zero. The best known upper bound is $O((n^2/\log^2{n}) (\log\log{n})^{O(1)})$ for real numbers \cite{GP14,Freund15,GS15,chan2018more} and $O(n^2(\log\log{n}/\log{n})^2)$ for integers \cite{BDP08}.} cannot be solved in $O(n^{2-\eps})$ time.
It is natural to ask: can we use these conjectures to rule out log-factor improvements for problems like LCS? 
And even more optimistically, one might hope to base the hardness of LCS on a more standard assumption like $\mathsf{P \neq NP}$.
Unfortunately, we can formally prove that these assumptions are not sharp enough to lead to any consequences for log-factor improvements, if only Turing reductions are used.
In Section~\ref{sec:disc} we prove the following theorem which also shows that an $O(f(n)/\log^c (f(n)))$ time algorithm for problem A cannot imply, via a fine-grained reduction, an $O(g(n)^{1-\eps})$ algorithm for problem $B$, unless $B$ is (unconditionally) solvable in $O(g(n)^{1-\delta})$ time. 

\begin{thm}[Informally]
\label{thm:logs}
If for some $c>0$ there is a fine-grained reduction proving that LCS is not in $O(n^2/\log^{c}n)$ time unless SETH fails, then SETH is false.
\end{thm}

Note that it also does not suffice to simply make SETH stronger by postulating a higher running time lower bound for CNF-SAT, since \emph{superpolynomial improvements are known} for this problem~\cite{PPSZ05,CIP06,DH09,AWY15}. Similarly, we cannot base a study of log-factor improvements on the APSP conjecture, since \emph{superlogarithmic improvements are known} for APSP~\cite{Wi14B}. (However, 3SUM could be a candidate to base higher lower bounds on, since only log-factor improvements are known~\cite{GP14,Freund15,GS15,BDP08}, see Section~\ref{sec:discussion} for a discussion.)

Thus, in a time when super-linear lower bounds for problems like LCS are far out of reach, and our only viable approach to obtaining such negative results is reductions-based, we are left with two options.
We could either leave the study of log-factor improvements in limbo, without a technology for proving negative results, or we could search for natural and convincing assumptions that are more fine-grained than SETH that could serve as the basis for the negative results we desire. 
Such assumptions were recently proposed by Abboud, Hansen, Vassilevska Williams and Williams \cite{AHVW16}.
The authors blame the hardness of shaving logs on the hardness of solving satisfiability on boolean formulas (Formula-SAT) faster than exhaustive search\footnote{In \cite{AHVW16} the authors focus on SAT on Branching Programs (BPs) rather than formulas, but due to standard transformations between BPs and formulas, the two problems are equivalent up to polynomial factors. Focusing on Formula-SAT will be crucial to the progress we make in this paper.}, by polynomial factors (which are log-factors in the runtime), a task for which there are well known ``circuit lower bound'' barriers in complexity theory.
They show that an $O(n^2/\log^{1000} n)$ algorithm for LCS would imply a major advance in circuit lower bounds. 
In the final section of this paper, we give a more detailed argument in favor of this approach.
Whether one should expect it to lead to \emph{tight} barriers, i.e. explaining the lack of $O(n^2/\log^{3} n)$ algorithms for LCS or any other natural problem, was completely unclear.
\paragraph*{The Machine Model}

We use the \emph{Word-RAM} model on words of size $\Theta(\log{n})$, where there is a set of operations on words that can be performed in time $\Oh(1)$. 
Most papers do not fix the concrete set of allowed operations, and instead refer to ``typical Boolean and arithmetic operations''. In this paper, we choose a set of operations $\mathcal P$ that is \emph{robust} with respect to changing the word size: For any operation $\circ \in \mathcal P$, given two words $a,b$ (of size $\Theta(\log{n})$) we can compute $a \circ b$ in time $(\log n)^{1+o(1)}$ on a Word RAM with word size $\Theta(\log \log n)$ and operation set $\mathcal P$. In other words, if we split $a,b$ into $\Theta(\log n / \log \log n)$ words of size $\Theta(\log \log n)$ then $a \circ b$ can still be computed very efficiently.

This robustness in particular holds for the following standard set of operations: initializing a cell with a constant, bitwise AND, OR, NOT, shift, addition, subtraction, multiplication, and division with remainder (since multiplication and division have near-linear time algorithms).

The results in this paper will get gradually weaker as we relax the restriction on near-linear time per operation to higher runtimes, however, even with this restriction, to the best of our knowledge this model captures all log shaving results in the literature (on the ``standard'' Word RAM model without fancy word operations). 



\paragraph*{Formula-SAT}
A boolean formula over $n$ input variables can be viewed as a tree in which every leaf is marked by an input variable or its negation and every internal node or \emph{gate} represents some basic boolean operation. 
Throughout this introduction we will only talk about \emph{deMorgan} formulas, in which every gate is from the set $\{ \wedge, \vee \}$.
 The \emph{size} of the formula is defined to be the number of leaves in the tree. 

In the Formula-SAT problem we are given a formula $F$ of size $s$ over $n$ inputs, and we have to decide whether there is an input $\{0,1\}^n$ that makes it output $1$.
A naive algorithm takes $O(2^n \cdot s)$ time, since evaluating the formula on some input takes $O(s)$ time.
Can we do better? 
We will call a SAT algorithm \emph{non-trivial}\footnote{Some works on SAT algorithms used this term for runtimes of the form $2^{n}\poly(s)/n^{\omega(1)}$. In our context, we need to be a bit more fine-grained. } if it has a runtime at most $O( \frac{2^n}{ n^\eps})$, for some $\eps>0$. 

It seems like a clever algorithm must look at the given formula $F$ and try to gain a speedup by analyzing it.
The more complicated $F$ can be, the harder the problem becomes.
Indeed, Dantsin and Hirsch \cite{DH09} survey dozens of algorithms for SAT on CNF formulas which exploit their structure. For $k$-CNF formulas of size $s$ there are $2^n s/2^{\Omega(n/k)}$ time algorithms (e.g.~\cite{PPSZ05}), and for general CNF formulas the bound is $2^n s/2^{\Omega(n/\log \Delta)}$ where $\Delta = s/n$ is the clause-to-variable ratio~\cite{CIP06,DH09,AWY15}.
The popular SETH \cite{IP01,CIP09} essentially says that this is close to optimal, and that there is no $2^{n} s / 2^{\Omega(n)}$ algorithm for CNF-SAT.
For arbitrary deMorgan formulas, the upper bounds are much worse. 
A FOCS'10 paper by Santhanam \cite{Santha10} and several recent improvements \cite{CK15,CKS14,CKKSZ15,KRT13,Tal15} solve Formula-SAT on formulas of size $s=n^{3-16\eps}$ in time $2^n s^{O(1)}/2^{n^\eps}$, which is non-trivial only  for $s =o(n^3)$, and going beyond cubic seems extremely difficult. 
This leads us to the first barrier which we will transform into a barrier for shaving logs.

\begin{hypo}
\label{hyp1}
There is no algorithm that can solve SAT on deMorgan formulas of size $s=n^{3+\Omega(1)}$ in $O(\frac{2^n}{n^\eps})$ time, for some $\eps>0$, in the Word-RAM model.
\end{hypo}

Perhaps the main reason to believe this hypothesis is that despite extensive algorithmic attacks on variants of SAT (perhaps the most extensively studied problem in computer science) over decades, none of the ideas that anyone has ever come up with seem sufficient to refute it.
Recent years have been particularly productive in non-trivial algorithms designed for special cases of Circuit-SAT \cite{Santha10,ST13,IMP12,CIP09,Wil14,BIS12,Chen15,IPS13,ILPS14,CS16,TSTT15,GKST16} (in addition to the algorithms for deMorgan formulas above) and this hypothesis still stands.

A well-known ``circuit lower bounds'' barrier seems to be in the way for refuting Hypothesis~\ref{hyp1}: can we find an explicit boolean function that cannot be computed by deMorgan formulas of cubic size?
Functions that require formulas of size $\Omega(n^{1.5})$ \cite{Subbo61} and $\Omega(n^{2})$ \cite{Khrap71} have been known since the 60's and 70's, respectively.
In the late 80's, Andreev \cite{Andreev87} proved an $\Omega(n^{2.5})$ which was later gradually improved to $\Omega(n^{2.55})$ by Nisan and Wigderson \cite{IN93} and to $\Omega(n^{2.63})$ by Paterson and Zwick \cite{PZ93} until H{\aa}stad proved his $n^{3-o(1)}$ lower bound in FOCS'93 \cite{Hastad98} (a recent result by Tal improves the $n^{o(1)}$ term \cite{Tal14}).
All these lower bound results use the ``random restrictions'' technique, first introduced in this context by Subbotovskaya in 1961 \cite{Subbo61}, and it is known that a substantially different approach must be taken in order to go beyond the cubic barrier. 
What does this have to do with Formula-SAT algorithms?
Interestingly, this same ``random restrictions'' technique was crucial to \emph{all} the non-trivial Formula-SAT algorithms mentioned above.
This is not a coincidence, but only one out of the many examples of the intimate connection between the task of designing non-trivial algorithms for SAT on a certain class $\mathcal{F}$ of formulas or circuits and the task of proving lower bounds against $\mathcal{F}$.
This connection is highlighted in many recent works and in several surveys \cite{Santha13,Oli13,Ryan_ICM14}.
The intuition is that both of these tasks seem to require identifying a strong structural property of functions in $\mathcal{F}$.
There is even a formal connection shown by Williams \cite{Wil13}, which in our context implies that solving Formula-SAT on formulas of size $O(n^{3.1})$ in $O(2^n/n^{10})$ time (which is only slightly stronger than refuting Hypothesis~\ref{hyp1}) is sufficient in order to prove that there is a function in the class ${\sf E}^{\sf NP}$ that cannot be computed by formulas of size $O(n^{3.1})$ (see \cite{AHVW16} for more details).
This consequence would be the first polynomial progress on the fundamental question of worst case formula lower bounds since H{\aa}stad's result.

\subsection{Our Results: New Reductions}

Many recent papers have reduced CNF-SAT to fundamental problems in ${\sf P}$ to prove SETH-based lower bounds (e.g. \cite{PW10,RV13,AVW14,AV14,Bring14,BI15,AVY15,ABHVZ16,CGR16,AVW16,BI16,MPS16,CDHL16}).
Abboud et al.~\cite{AHVW16} show that even SAT on formulas, circuits, and more, can be efficiently reduced to combinatorial problems in ${\sf P}$.
In particular, they show that Formula-SAT on formulas of size $s$ over $n$ inputs can be reduced to an instance of LCS on sequences of length $N=O(2^{n/2} \cdot s^{1000})$.
This acts as a barrier for shaving logs as follows.
A hypothetical $O(N^2/\log^{c}{N})$ time algorithm for LCS can be turned into an 
$$ n^{1+o(1)} \cdot (2^{n/2}\cdot s^{1000})^2/ (\log{2^{\Omega(n)}})^c = O(2^n \cdot s^{2000} / n^{c-1})$$ time algorithm for Formula-SAT, which for a large enough $c \geq 2001$ would refute Hypothesis~\ref{hyp1}.
The first $n^{1+o(1)}$ factor in the runtime comes from the jump from $n$ to $N=2^n$ and our Word-RAM machine model: whenever the LCS algorithm wants to perform a unit-cost operation on words of size $\Theta(\log{N})$ (this is much more than the word size of our SAT algorithm which is only $\Theta(\log{n}) = \Theta(\log\log{N})$), the SAT algorithm can simulate it in  $(\log{N})^{1+o(1)} = n^{1+o(1)}$ time in the Word-RAM model with words of size $\Theta(\log{n})$.

Our main result is a \emph{much} more efficient reduction to LCS. 
For large but constant size alphabets, we get a near-linear dependence on the formula size, reducing the $s^{1000}$ factor to just $s^{1+o(1)}$.

\begin{thm}
\label{thm:lcs}
Formula-SAT on formulas of size $s$ on $n$ inputs can be reduced to an instance of LCS on two sequences over an alphabet of size $\sigma$ of length $N= 2^{n/2} \cdot s^{1+O(1/\log\log\sigma)}$, in $O(N)$ time.
\end{thm}


Thus, if LCS on sequences of length $N$ and alphabet of size $\omega(1)$ can be solved in $O(N^2/\log^c{N})$ time, then Formula-SAT can be solved in $2^n \cdot \frac{ s^{2+o(1)}}{n^{c}} \cdot n^{1+o(1)}$ time.
Recall that the known upper bound for LCS is $O(n^2/\log^cn)$ for any constant alphabet size, with $c=2$, and we can now report that the barrier of cubic formulas stands in the way of improving it to $c>7$ (see Corollary~\ref{cor:lcs} below).
%

The novelty in the proof of Theorem~\ref{thm:lcs} over \cite{AHVW16} is discussed in Section~\ref{sec:technical}. As an alternative to Theorem~\ref{thm:lcs}, in Section~\ref{sec:lcssimple} we present another reduction to LCS which is much simpler than all previously known reductions, but uses a larger alphabet.



\paragraph*{Fr\'echet Distance}

An important primitive in computational geometry is to judge how similar are two basic geometric objects, such as polygonal curves, represented as sequences of points in $d$-dimensional Euclidean space. 
Such curves are ubiquitous, since they arise naturally as trajectory data of moving objects, or as time-series data of stock prices and other measures. 
The most popular similarity measure for curves in computational geometry is the \Fr distance, also known as dog-leash-distance.
For formal definitions see Section~\ref{sec:frechet}. 
The \Fr distance has found many applications (see, e.g.,~\cite{MunichP99,BrakatsoulasPSW05,BuchinBGLL11}) and developed to a rich field of research with many generalizations and variants (see, e.g.,~\cite{AltB10,AronovHPKW06,AltKW04,DriemelHPW12,ChambersETAL10,
Wenk2010geodesic,BuchinBW09,DriemelHP13,MaheshwariSSZ11,Indyk02}).

This distance measure comes in two variants: the continuous and the discrete.
A classic algorithm by Alt and Godau~\cite{AltG95,Godau91} computes the continuous \Fr distance in time $\Oh(n^2 \log n)$ for two given curves with $n$ vertices. The fastest known algorithm runs in time $\Oh(n^2 (\log \log n)^2)$ (on the Word RAM)~\cite{BuchinBMM14}. If we only want to decide whether the \Fr distance is at most a given value $\delta$, this algorithm runs in time $\Oh(n^2 (\log \log n)^2 / \log n)$.
For the discrete \Fr distance, the original algorithm has running time $\Oh(n^2)$~\cite{EiterM94}, which was improved to $\Oh(n^2 \log \log n / \log n)$ by Agarwal et al.~\cite{AgarwalBAKS13}. Their algorithm runs in time $\Oh(n^2 \log \log n / \log^2 n)$ for the decision version. 
It is known that both versions of the \Fr distance are SETH-hard~\cite{Bring14}. However, this does not rule out log factor improvements. In particular, no reduction from versions of SETH on formulas or branching programs is known.

In this paper we focus on the \emph{decision version of the discrete \Fr distance} (which we simply call ``\Fr distance'' from now on). 
We show that \Fr distance suffers from the same barriers for shaving logs like LCS.
In particular, this reduction allows us to base the usual $\Omega(n^{2-\eps})$ lower bound on a weaker assumption than SETH, such as NC-SETH (see the discussion in \cite{AHVW16}). This is the first NC-SETH hardness for a problem that does not admit alignment gadgets (as in \cite{BK15}).

\begin{thm}
\label{thm:frechet}
Formula-SAT on formulas of size $s$ on $n$ inputs can be reduced to an instance of the \Fr distance on two curves of length $N=\Oh(2^{n/2} \cdot s)$, in $O(N)$ time.
\end{thm}

\paragraph*{Regular Expression Pattern Matching}

Our final example is the fundamental \emph{Regular Expression Pattern Matching} problem: Decide whether a given regular expression of length $m$ matches a substring of a text of length $n$. Again, there is a classical $O(nm)$ algorithm \cite{Tho68}, and the applicability and interest in this problem resulted in algorithms shaving log factors; the first one by Myers \cite{Mye92} was improved by Bille and Thorup \cite{BT09} to time $O(mn/\log^{1.5}{n})$.
Recently, Backurs and Indyk proved an $n^{2-o(1)}$ SETH lower bound \cite{BI16}, and performed an impressive study of the exact time complexity of the problem with respect to the complexity of the regular expression. This study was essentially completed by Bringmann, Gr{\o}nlund, and Larsen \cite{BGL16}, up to $n^{o(1)}$ factors.
In Section~\ref{sec:patternmatch} we show that this problem is also capable of efficiently simulating formulas and thus has the same barriers as LCS and \Fr distance.

\begin{thm}
\label{thm:patternmatch}
Formula-SAT on formulas of size $s$ on $n$ inputs can be reduced to an instance of Regular Expression Pattern Matching on text and pattern of length $N=\Oh(2^{n/2} \cdot s \log s)$ over a constant size alphabet, in $O(N)$ time.
\end{thm}

\paragraph*{Consequences of the Cubic Formula Barrier}

We believe that SAT on formulas can be tightly connected to many other natural problems in P. 
As we discuss in the next section, such reductions seem to require problem-specific engineering and are left for future work.
The main point of this paper is to demonstrate the possibility of basing such ultra fine-grained lower bounds on one common barrier.
Our conditional lower bounds are summarized in the following corollary, which shows that current log-shaving algorithms are very close to the well-known barrier from complexity theory of cubic formula lower bounds.

\begin{cor}
\label{cor:lcs}
For all $\eps>0$, solving any of the following problems in $O(n^2/\log^{7+\eps}{n})$ time refutes Hypothesis~\ref{hyp1}, and solving them in $O(n^2/\log^{17+\eps}{n})$ time implies that ${\sf E}^{\sf NP}$ cannot be computed by non-uniform formulas of cubic size:

\begin{itemize}
\item LCS over alphabets of size~$\omega(1)$
\item The \Fr distance on two curves in the plane
\item Regular Expression Pattern Matching over constant size alphabets.
\end{itemize}

\end{cor}

The main reason that our lower bounds above are not tight (the gap between $2$ and $7$) is that we need to start from SAT on \emph{cubic} size formulas rather than linear size ones, due to the fact that clever algorithms do exist for smaller formulas. 
We remark that throughout the paper we will work with a class of formulas we call $\mathcal{F}_1$ (see Section~\ref{sec:simple}), also known as bipartite formulas, that are more powerful than deMorgan formulas yet our reduction to LCS can support them as well.
This makes our results stronger, since $\mathcal{F}_1$-Formula-SAT could be a harder problem than SAT on deMorgan formulas.
In fact, in an earlier version of the paper we had suggested the hypothesis that $\mathcal{F}_1$-Formula-SAT does not have non-trivial algorithms even on \emph{linear size} formulas. 
This stronger hypothesis would give higher lower bounds.
However, Avishay Tal (personal communication) told us about such a non-trivial algorithm for formulas of size up to $n^{2-\Omega(1)}$ using tools from quantum query complexity.
We are optimistic that one could borrow such ideas or the ``random restrictions'' technique from SAT algorithms in order to shave more logs for combinatorial problems such as LCS.
This is an intriguing direction for future work.

\section{Technical Overview and the Reduction to LCS}
\label{sec:technical}


All the reductions from SAT to problems in P mentioned above start with a split-and-list reduction to some ``pair finding" problem.
In the SETH lower bounds, CNF-SAT is reduced to the Orthogonal-Vectors problem of finding a pair $a \in A, b \in B, A,B \subseteq \{0,1\}^d$ that are orthogonal \cite{Wil05}.
When starting from Formula-SAT, we get a more complex pair-finding problem.
In Section~\ref{sec:simple} we show a simple reduction from SAT on formulas from the class $\mathcal{F}_1$ (which contains deMorgan formulas) to the following problem.

\begin{defn}[\textsc{Formula-Pair} Problem]
Given a deMorgan formula over $2m$ variables $F=F(x_1,\ldots,x_m,y_1,\ldots,y_m)$  (each appearing once in $F$), and two sets of vectors $A,B \subseteq \{0,1\}^{m}$ of size $n$, decide if there is a pair $a \in A, b \in B$ such that $F(a_1,\ldots,a_m,b_1,\ldots,b_m)=\true$.
\end{defn}


In Section~\ref{sec:simple} we show a Four-Russians type algorithm that solves \textsc{Formula-Pair} in $O(n^2m/\log^2{n})$ time, and
even when $m=|F|=(\log{n})^{1+o(1)}$ no $O(n^2/\log^{1+\eps}n)$ upper bound is known. By our reduction, such an upper bound would imply a non-trivial algorithm for SAT on formulas from $\mathcal{F}_1$.
Moreover, Hypothesis~\ref{hyp1} implies that we cannot solve \textsc{Formula-Pair} in $O(n^2/\log^{\eps}{n})$ time, for $m=(\log{n})^{3+\Omega(1)}$.
In the next sections, we reduce \textsc{Formula-Pair} to LCS, from which Theorem~\ref{thm:lcs} follows.
A simpler reduction using much larger alphabet size can be found in Section~\ref{sec:lcssimple}.

\begin{thm}
\label{thm:lcspair}
\textsc{Formula-Pair} on formulas of size $s$ and lists of size $n$ can be reduced to an instance of LCS on two strings over alphabet of size $\sigma \geq 2$ of length $O(n \cdot s^{1+O(1/\log\log \sigma)})$, in linear time. 
\end{thm}

The reduction constructs strings $x,y$ and a number $\rho$ such that $\LCS(x,y) \ge \rho$ holds if and only if the given \SuccPair instance $(\F,A,B)$ is satisfiable. 
The approach is similar to the reductions from Orthogonal-Vectors to sequence alignment problems (e.g.~\cite{AVW14,Bring14,BI15,ABV15,BK15}).
The big difference is that our formula $F$ can be much more complicated than a CNF, and so we will need more powerful gadgets.
Sequence gadgets that are able to simulate the evaluation of deMorgan formulas were (implicitly) constructed in \cite{AHVW16} with a recursive approach.
Our main contribution is an \emph{extremely efficient} implementation of such gadgets with LCS.

The main part of the reduction is to construct \emph{gate gadgets}: for any vectors $a,b \in \{0,1\}^m$ and any gate $g$ of $\F$, we construct strings $x(g,a)$ and $y(g,b)$ whose LCS determines whether gate $g$ evaluates to true for input $(a,b)$ to $\F$ (see \secref{lcsgates}).
Once we have this, to find a pair of vectors $a \in A, b \in B$ satisfying $\F$, we combine the strings $x(r,a),y(r,b)$, constructed for the root $r$ of $\F$, using  a known construction of so-called alignment gadgets~\cite{ABV15,BK15} from previous work (see \secref{lcsalignmentgadgets}). 

Let us quickly explain how \cite{AHVW16} constructed gate gadgets and the main ideas that go into our new construction.
There are two kinds of gadgets, corresponding to the two types of gates in $F$: AND and OR gates.
Since the AND gadgets will be relatively simple, let us consider the OR gadgets.
Fix two inputs $a,b$, and let $g=(g_1 \vee g_2)$ be an OR gate, and assume that we already constructed gate gadgets for $g_1,g_2$, namely $x_1 = x(g_1,a), y_1=y(g_1,b), x_2 = x(g_2,a), y_2=y(g_2,b)$ so that for $i \in \{1,2\}$ we have that $\LCS(x_i,y_i)$ is large if the gate $g_i$ outputs true on input $(a,b)$, and it is smaller otherwise.
In \cite{AHVW16}, these gadgets were combined as follows. 
Let $\beta$ be an upper bound on the total length of the gadgets $x_i,y_i$. 
We add a carefully chosen padding of $0$'s and $1$'s, so that any optimal matching of the two strings will have to match \emph{either} $x_1,y_1$ or $x_2,y_2$ but not both. 

\begin{alignat*}{6}
  x \; &:= \quad &&0^{4\beta} \;  x_1 \; 1^{\beta}   \; x_2 \;  0^{4\beta}  \\
  y \; &:= \quad &&\;  y_1 \; 1^{\beta} 0^{4\beta} 1^\beta  \; y_2 \; 
\end{alignat*}

One then argues that, in any optimal LCS matching of $x,y$, the $0^{4\beta}$ block of $y$ must be matched either left or right. If it's matched left, then the total score will be equal to $4\beta+\beta+LCS(x_2,y_2)$ while if it's matched right, we will get $4\beta+\beta+LCS(x_1,y_1)$.
Thus, $LCS(x,y)$ is determined by the OR of $g_1,g_2$.
The blowup of this construction is a multiplicative factor of $11$ with every level of the formula, and the length of the gadget of the root will end up roughly $11^{\depth(F)}$. 
To obtain our tight lower bounds, we will need to decrease this blowup to $1+\eps_\sigma$ at every level, where $\eps_\sigma$ goes to 0 when the alphabet size $\sigma$ tends to infinity.
With the above construction, decreasing the length of the padding will allow the optimal LCS matching to cheat, e.g.\ by matching $y_1$ to both $x_1$ and $x_2$, and no longer corresponding to the OR of $g_1,g_2$.

Our first trick is an ultra-efficient OR gadget in case we are allowed unbounded alphabet size.
We take $x_1,y_1$ and transform all their letters into a new alphabet $\Sigma^{g_1}$, and we take $x_2,y_2$ and transform their letters into a disjoint alphabet $\Sigma^{g_2}$.
Then our OR gadget does not require any padding at all:
\begin{alignat*}{6}
  x \; &:= \quad && \;  x_1 \;   \; x_2 \;   \\
  y \; &:= \quad &&\;  y_2 \;   \; y_1 \; 
\end{alignat*}
The crossing structure of this construction means that any LCS matching that matches letters from $x_1,y_1$ cannot also match letters from $x_2,y_2$, and vice versa, while the disjoint alphabets make sure that there can be no matches between $x_1,y_2$ or $x_2,y_1$.
With such gadgets we can encode a formula of size $s$ with $O(s)$ letters, for details see Section~\ref{sec:lcssimple}.

But how would such an idea work for \emph{constant size alphabets}?
Once we allow $x_1$ and $y_2$ to share even a single letter, this argument breaks.
Natural attempts to simulate this construction with smaller alphabets, e.g.\ by replacing each letter with a random sequence, do not seem to work, and we do not know how to construct such an OR gadget with a smaller alphabet \emph{in a black box way}.
The major part of our proof will be a careful examination of the formula and the sub-gadgets $g_1,g_2$ in order to reduce the alphabet size to a large enough constant, while using padding that is only $1+\eps_\sigma$ times the length of the sub-gadgets.
We achieve this by combining this crossing gadget with a small padding that will reuse letters from alphabets that were used much deeper in the formula, and we will argue that the noise we get from recycling letters is dominated by our paddings, in any optimal matching.

We remark that the reduction of \cite{AHVW16} can be implemented in a generic way with any problem that admits \emph{alignment gadgets} as defined in \cite{BK15}, giving formula-gadgets of size $s^{O(1)}$.
The list of such problems includes LCS and Edit-Distance on binary strings.
However, to get gadgets of length $s^{1+o(1)}$ it seems that problem-specific reductions are necessary.
A big open question left by our work is to find the most efficient reduction from Formula-SAT to Edit-Distance.
A very efficient OR gadget, even if the alphabet is unbounded, might be (provably) impossible.
Can we use this intuition to shave more log factors for Edit-Distance?

Fr\'echet Distance falls outside the alignment gadgets framework of \cite{BK15} and no reduction from Formula-SAT was known before.
In Section~\ref{sec:frechet} we prove such a reduction by a significant boosting of the SETH-lower bound construction of \cite{Bring14}.
In order to implement recursive AND/OR gadgets, our new proof utilizes the geometry of the curves, in contrast to \cite{Bring14} which only used ten different points in the plane. 

\medskip
In the remainder of this section we present the details of the reduction to LCS. Some missing proofs can be found in Section~\ref{sec:lcsproofs}.

\subsection{Implementing Gates}
\label{sec:lcsgates}

Fix vectors $a,b \in \{0,1\}^m$ (where $2m$ is the number of inputs to $\F$). 
In this section we prove the following lemma which demonstrates our main construction.

\begin{lem} \label{lem:lcsgatesconstr}
  For any sufficiently large $\sigma > 0$ let $\tau = (\log \sigma)^{1/4}$.
  We can inductively construct, for each gate $g$ of $\F$, strings $x(g)=x(g,a)$ and $y(g)=y(g,b)$ over alphabet size $5 \sigma^{2}$ and a number $\rho(g)$ such that for $L(g) := \LCS(x(g),y(g))$ we have (1) $L(g) \le \rho(g)$ and (2) $L(g) = \rho(g)$ if and only if gate $g$ evaluates to true on input $(a,b)$ to $\F$. Moreover, we have $|x(g)| = |y(g)| = n(g) \le 6 \tau \cdot |\F_g| (1+7/\tau)^{\depth(\F_g)}$, where $\F_g$ is the subformula of $\F$ below $g$.
\end{lem}

In this construction, we use disjoint size-5 alphabets $\Sigma_1,\ldots,\Sigma_{\sigma^2}$, determining the total alphabet size as $5 \sigma^2$. Each gate $g$ is assigned an alphabet $\Sigma_{f(g)}$. We fix the function $f$ later. 

In the following, consider any gate $g$ of $\F$, and write the gate alphabet as $\Sigma_{f(g)} = \{0,1,2,3,4\}$. For readability, we write $x = x(g)$ and similarly define $y,n,L,\rho$.
If $g$ has fanin 2, write $g_1,g_2$ for the children of $g$. Moreover, let $x_1 = x(g_1)$ and similarly define $y_1,n_1,L_1,\rho_1$ and $x_2,y_2,n_2,L_2,\rho_2$. 

\paragraph*{Input Gate}
The base case is an input bit $a_i$ to $\F$ (input bits $b_j$ are symmetric).
Interpreting $a_i$ as a string of length~1 over alphabet $\{0,1\}$, note that $\LCS(a_i,1) = a_i$. Hence, the strings $x = a_i$ and $y = 1$, with $n=\rho=1$, trivially simulate the input bit $a_i$.


\paragraph*{AND Gates}
Consider an AND gate $g$ and let $\beta := \lceil (n_1 + n_2)/\tau^2 \rceil$. We construct strings $x,y$ as
\begin{alignat*}{6}
  x \; &:= \quad &&x_1 \; & &0^{\beta}  \; 1^{\beta} & &  \; x_2 \\
  y \; &:= \quad &&y_1 \; & &0^{\beta}  \; 1^{\beta} & &  \; y_2
\end{alignat*}
\begin{lem} \label{lem:AND}
  If $\LCS(x_2,y_1), \LCS(x_1,y_2) \le \beta/4$ and the symbols $0,1$ appear at most $\beta/16$ times in each of $x_1,x_2,y_1$, and $y_2$, then we have $L = \LCS(x,y) = 2\beta + L_1 + L_2$.
\end{lem}

Later we will choose the gate alphabets $\Sigma_{f(g)}$ such that the precondition of the above lemma is satisfied. Setting $\rho := 2\beta + \rho_1 + \rho_2$ we thus inductively obtain (1) $L \le \rho$ and (2) $L = \rho$ if and only if $g_1$ and $g_2$ both evaluate to true. Thus, we correctly simulated the AND gate $g$. It remains to prove the lemma. 

\begin{proof}
  Clearly, we have $L \ge \LCS(x_1,y_1) + \LCS(0^{\beta},0^{\beta}) + \LCS(1^{\beta},1^{\beta}) + \LCS(x_2,y_2) = 2\beta + L_1 + L_2$. For the other direction, consider any LCS $z$ of $x,y$. If $z$ does not match any symbol of the left half of $x$, $x_1 0^{\beta}$, with any symbol of the right half of $y$, $1^{\beta} y_2$, and it does not match any symbol of the right half of $x$, $1^{\beta} x_2$, with any symbol of the left half of $y$, $y_1 0^{\beta}$, then we can split both strings in the middle and obtain 

  $$ L = |z| \le \LCS(x_1 0^{\beta}, y_1 0^{\beta}) + \LCS(1^{\beta} x_2, 1^{\beta} y_2) .$$ 

  Greedy suffix/prefix matching now yields 
  $$ L \le \big( \LCS(x_1, y_1) + \beta \big) + \big(\beta + \LCS(x_2,y_2) \big) = 2\beta + L_1 + L_2. $$

  In the remaining case, there is a matching from some left half to some right half. By symmetry, we can assume that there is a matching from the left half of $x$ to the right half of $y$. We can moreover assume that $z$ matches a symbol of $x_1$ with a symbol of $1^{\beta} y_2$, since the case that $z$ matches a symbol of $y_2$ with a symbol of $x_1 0^{\beta}$ is symmetric. 
  Now no symbol in $0^{\beta}$ in $x$ can be matched with a symbol in $0^{\beta}$ in $y$. We obtain a rough upper bound on $L = |z|$ by summing up the LCS length of all remaining $4\cdot 4 - 1=15$ pairs of a part $x' \in \{x_1, 0^{\beta}, 1^{\beta}, x_2\}$ in $x$ and a part $y' \in \{y_1, 0^{\beta}, 1^{\beta}, y_2\}$ in $y$. This yields $L \le L_1 + L_2 + \beta + 2 \cdot \beta/4 + 8 \cdot \beta/16 = 2\beta + L_1 + L_2$, finishing the proof.
\end{proof}

\paragraph*{OR Gates}
Consider an OR gate $g$ and again let $\beta := \lceil (n_1 + n_2)/\tau^2 \rceil$. We first make the LCS target values equal by adding $4^{|\rho_1-\rho_2|}$ to the shorter of $x_2/y_2$ and $x_1/y_1$, i.e., we set $x'_1 := 4^{\max\{0,\rho_2-\rho_1\}} x_1$ and similarly $y'_1 := 4^{\max\{0,\rho_2-\rho_1\}} y_1$, $x'_2 := 4^{\max\{0,\rho_1-\rho_2\}} x_2$, $y'_2 := 4^{\max\{0,\rho_1-\rho_2\}} y_2$. Note that the resulting strings satisfy $L'_1 := \LCS(x'_1,y'_1) \le \rho' := \max\{\rho_1,\rho_2\}$ and $L'_1 = \rho'$ if and only if $g_1$ evaluates to true, and similarly $L'_2 := \LCS(x'_2,y'_2) \le \rho'$ and $L'_2 = \rho'$ if and only if $g_2$ evaluates to true.
We construct the strings $x,y$ as
\begin{alignat*}{6}
  x \; &:= \quad &&0^{\beta} 1^\beta\; && x'_1 \; & &2^{\beta} 3^\beta  & &  \; x'_2 \; && 0^\beta 1^{\beta} \\
  y \; &:= \quad &&2^{\beta} 3^\beta\; && y'_2 \; & &0^{\beta} 1^\beta  & &  \; y'_1 \; && 2^{\beta} 3^\beta
\end{alignat*}
\begin{lem} \label{lem:OR}
  If $\LCS(x_2,y_1), \LCS(x_1,y_2) \le \beta/8$ and the symbols $0,1,2,3$ appear at most $\beta/48$ times in each of $x_1,x_2,y_1$, and $y_2$, then $L = \LCS(x,y) = 4\beta + \max\{L'_1, L'_2\}$.
\end{lem}

Later we will choose the gate alphabets $\Sigma_{f(g)}$ such that the precondition of the above lemma is satisfied. Setting $\rho := 4\beta + \rho' = 4\beta + \max\{\rho_1,\rho_2\}$ we thus inductively obtain (1) $L \le \rho$ and (2) $L = \rho$ if and only if at least one of $g_1$ and $g_2$ evaluates to true, so we correctly simulated the OR gate $g$. The proof of the Lemma is in Section~\ref{sec:lcsproofs}.

\paragraph*{Analyzing the Length}
Note that the above constructions inductively yields strings $x(g),y(g)$ simulating each gate $g$. We inductively prove bounds for $n(g)$ and $\rho(g)$. See Section~\ref{sec:lcsproofs}.

\begin{lem}
\label{lem:length}
  We have $n(g) \le 6 \tau \cdot |\F_g| (1+7/\tau)^{\depth(\F_g)}$ and $\rho(g) \le 6 |\F_g| (1+7/\tau)^{\depth(\F_g)}$ for any gate $g$, where $\F_g$ is the subformula of $\F$ below $g$.
\end{lem}

\paragraph*{Fixing the Gate Alphabets}
Now we fix the gate alphabet $\Sigma_{f(g)}$ for any gate $g$. 
Again let $\Sigma_{(i,j)}$, $i,j \in [\sigma]$, be disjoint alphabets of size 5, and let $\Sigma := \bigcup_{i,j} \Sigma_{(i,j)}$. 
For any gate $g$ of $\F$, we call its distance to the root the \emph{height} $h(g)$. For any $h$, order the gates with height $h$ from left to right, and let $\iota(g)$ be the index of gate $g$ in this order, for any gate $g$ with height $h$. Note that $(h(g),\iota(g))$ is a unique identifier of gate~$g$. We define $f(g) := (h(g) \bmod \sigma, \iota(g) \bmod \sigma)$, i.e., we set the gate alphabet of $g$ to $\Sigma_{f(g)} = \Sigma_{(h(g) \bmod \sigma, \iota(g) \bmod \sigma)}$.
Note that the overall alphabet $\Sigma$ has size $5 \sigma^2$. Recall that we set $\tau := (\log \sigma)^{1/4}$.

It remains to show that the preconditions of \lemrefs{AND}{OR} are satisfied. Specifically, consider a gate $g$ with children $g_1,g_2$. As before, let $x,y,n$ be the strings and string length constructed for gate $g$, and let $x_i,y_i,n_i$ be the corresponding objects for $g_i$, $i \in \{1,2\}$. We need to show: 
\begin{enumerate}[label=(\arabic*)]
  \item $\LCS(x_2,y_1), \LCS(x_1,y_2) \le (n_1+n_2)/(8 \tau^2)$, and 
  \item each symbol $c \in \Sigma_{f(g)}$ appears at most $(n_1+n_2)/(48 \tau^2)$ times in each of $x_1,x_2,y_1$, and $y_2$. 
\end{enumerate}

We call a gate $g'$ in the subformula $\F_g$ \emph{$d$-deep} if $h(g') \ge h(g) + d$, and \emph{$d$-shallow} otherwise.
For each symbol $c$ in $x$ or $y$ we can trace our construction to find the gate $g'$ in $\F_g$ at which we introduced $c$ to $x$ or $y$. In other words, each symbol in $x,y$ \emph{stems} from some gate $g'$ below~$g$. 

First consider (2). 
Observe that all symbols in $x,y$ stemming from $\sigma$-shallow gates do not belong to the gate alphabet $\Sigma_{f(g)}$, since the function $f(g')$ has $(h(g') \bmod \sigma)$ as the first component, which repeats only every $\sigma$ levels. Thus, if a symbol $c \in \Sigma_{f(g)}$ occurs in $x_i$ or $y_i$, then this occurence stems from a $\sigma$-deep gate. We now argue that only few symbols in $x,y$ stem from deep gates. For any $d > 0$, let $N_d$ be the number of symbols in $x$ (or, equivalently, $y$) steming from $d$-deep gates. Note that $N_d$ is equal to the total string length $\sum n(g')$, summed over all gates $g'$ in $\F_g$ with height $h(g') = h(g)+d$. Observe that our construction increases the string lengths in each step by at least a factor $1+1/\tau^2$, i.e., $N_{d} \ge (1+1/\tau^2) N_{d+1}$ holds for any $d$. It follows that $N_{\sigma} \le N_{1} / (1+1/\tau^2)^{\sigma-1} = (n_1+n_2) / (1+1/\tau^2)^{\sigma-1}$. Hence, each symbol in $\Sigma_{f(g)}$ appears at most $(n_1 + n_2) / (1+1/\tau^2)^{\sigma-1}$ times in each of $x_1,x_2,y_1,y_2$. Since $\tau = (\log \sigma)^{1/4}$, we have $(1+1/\tau^2)^{\sigma-1} = 2^{\Omega(\sigma / \sqrt{\log \sigma})} \ge 48 \sqrt{\log \sigma} = 48 \tau^2$ for sufficiently large $\sigma$. This proves (2).

For (1), remove all $\log(\sigma)$-deep symbols from $x_1$ and $y_2$ to obtain strings $x'_1,y'_2$. Note that we removed exactly $N_{\log \sigma}$ symbols from each of $x_1,y_2$. This yields $\LCS(x_1,y_2) \le 2 N_{\log \sigma} + \LCS(x'_1,y'_2)$. For $x'_1,y'_2$, we claim that any $\log(\sigma)$-shallow gates $g'_1 \ne g'_2$ in $\F_g$ have disjoint alphabets $\Sigma_{f(g'_1)},\Sigma_{f(g'_2)}$. Indeed, if $h(g'_1) \ne h(g'_2)$ then since the first component $(h(g') \bmod \sigma)$ of $f(g')$ repeats only every $\sigma$ levels we have $f(g'_1) \ne f(g'_2)$. If $h(g'_1) = h(g'_2) =: h$, then note that each gate $g'$ in height $h$ has a unique label $\iota(g') \bmod \sigma$, since there are $\sigma$ such labels and there are at most $2^{h-h(g)} < \sigma$ gates with height $h$ in $\F_g$. Hence, $x'_1$ and $y'_2$ use disjoint alphabets, and we obtain $\LCS(x'_1,y'_2) = 0$. Thus, $\LCS(x_1,y_2) \le 2 N_{\log \sigma}$.
As above, we bound $N_{\log \sigma} \le (n_1 + n_2) / (1+1/\tau^2)^{\log \sigma - 1}$, so that $\LCS(x_1,y_2) \le 2 (n_1 + n_2) / (1+1/\tau^2)^{\log \sigma - 1}$. Since $\tau = (\log \sigma)^{1/4}$, we have $(1+1/\tau^2)^{\log \sigma - 1}/2 = 2^{\Omega(\sqrt{\log \sigma})} \ge 8 \sqrt{\log  \sigma} = 8 \tau^2$ for sufficiently large $\sigma$. This yields (1), since the strings $x_2,y_1$ are symmetric.
This finishes the proof of \lemref{lcsgatesconstr}.

\paragraph*{Finalizing the Proof}
Let us sketch how we complete the proof of Theorem~\ref{thm:lcspair}. The full details are in Section~\ref{sec:lcsalignmentgadgets}.
First, for all vectors $a\in A, b \in B$ we construct gate gadgets for the output gate of the formula, i.e.\ \emph{formula gadgets}, by invoking \lemref{lcsgatesconstr}.
Then we combine all these gadgets by applying a standard alignment gadget~\cite{ABV15,BK15} to get our final sequences of length $\Oh\big(n\tau |\F| (1+7/\tau)^{\depth(\F)}\big)$ and with alphabet of size $O(\sigma^2)$.
The LCS of the final sequence will be determined by the existence of a satisfying pair.
Since a priori the depth of $\F$ could be as large as $|\F|$, the factor $(1+7/\tau)^{\depth(\F)}$ in our length bound is not yet satisfactory.
Thus, as a preprocessing before the above construction, we decrease the depth of $\F$ using a depth-reduction result of Bonet and Buss \cite{Spira71,BB94}: for all $k \ge 2$ there is an equivalent formula $\F'$ with depth at most $(3 k \ln 2) \log |\F|$ and size $|\F'| \le |\F|^{1 + 1/(1 + \log(k-1))}$.
Choosing the parameters correctly, we get final sequences of length $\Oh\big(n |\F|^{1 + \Oh(1/\log \log \sigma)}\big)$.

	\section{On the Limitations of Fine-Grained Reductions}
	\label{sec:disc}
With the increasingly complex web of reductions and conjectures used in the ``Hardness in P" research, one might oppose to our use of nonstandard assumptions.
Why can't we base the hardness of shaving logs on one of the more established assumptions such as SETH, or even better, on $\mathsf{P \neq NP}$?
We conclude the paper with a proof that such results are not possible if one is restricted to fine-grained reductions, which is essentially the only tool we have in this line of research.

Let $A$ be a problem with best known upper bound of $T_A(n)$ on inputs of size $n$,
and let $B$ be a problem with best known upper bound of $T_B(n)$ on inputs of size $n$. 
Throughout this section we assume that these runtime are non-decreasing functions, such as $2^n$ or $n^2$.
A fine-grained reduction from ``solving $A$ in time $T_A(n)/g(n)$" to ``solving $B$ in time $T_B(n)/f(n)$" proves that improving $T_B(n)$ to $T_B(n)/f(n)$  improves $T_A$ to $T_A(n)/g(n)$.
Formally, it is an algorithm $X$ that solves $A$ and it is allowed to call an oracle for problem $B$, as long as the following bound holds.
Let $n_i$ be the size of the instance in the $i^{th}$ call to problem $B$ that our algorithm performs, where $i \leq t$ for some value $t$, and let $T_X(n)$ be the runtime of $X$ excluding the time it takes to answer all the instances of problem $B$. 
It must be that $T_X(n)+\sum_{i=1}^t T_B(n_i)/f(n_i) \leq T_A(n)/g(n)$.
This is a natural adaptation of the definition of fine-grained reductions from previous works, where the improvements were restricted to be by polynomial factors.

We can now give a formal version of Theorem~\ref{thm:logs} from the introduction. Note that $k$-SAT on $n$ variables and $m$ clauses can be solved in time $\textup{poly}(n,m) 2^n$.

\begin{thm}
\label{thm:logsformal}
If for some $c, \eps>0$ and all $k \ge 2$ there is a fine-grained reduction from solving $k$-SAT in time $\poly(n,m) 2^n / 2^{\eps n}$ to solving LCS in time $O(n^2/\log^{c}n)$, then SETH is false.
\end{thm}

\begin{proof}
Assume there was a fine-grained reduction from $k$-SAT to LCS as above. 
This means that there is an algorithm $X$ for $k$-SAT that makes $t$ calls to LCS with instances of size $n_1,\ldots,n_t$ such that:
$$
T_X(n) + \sum_{i=1}^t n_i^2 / \log^c n_i = O(\poly(n,m) 2^n / 2^{\eps n})
$$

But then consider algorithm $X'$ which simulates $X$ and whenever $X$ makes a call to the LCS oracle with an instance of size $n_i$, our algorithm will execute the known quadratic time solution for LCS that takes $O(n_i^2)$ time.
Let $n_{max}$ be the size of the largest instance we call, and note that $n_{max} < 2^n$. 
Simple calculations show that $X'$ solves $k$-SAT and has a running time of 
$$
T_X(n) + \sum_{i=1}^t n_i^2  =  O\big(\poly(n,m) 2^n / 2^{\eps n}\big) \cdot \log^c {n_{max}} = O(\poly(n,m) 2^n / 2^{\eps n})
$$
for all $k$, refuting SETH.
\end{proof}

\medskip
\paragraph*{Acknowledgements} 
We are grateful to Avishay Tal for telling us about his algorithm for SAT on bipartite formulas, and for very helpful discussions.
We also thank Mohan Paturi, Rahul Santhanam, Srikanth Srinivasan, and Ryan Williams for answering our questions about the state of the art of Formula-SAT algorithms, and Arturs Backurs, Piotr Indyk, Mikkel Thorup, and Virginia Vassilevska Williams for helpful discussions regarding regular expressions. 
We also thank an anonymous reviewer for ideas leading to shaving off a second log-factor for \textsc{Formula-Pair}, and other reviewers for helpful suggestions.

	\bibliographystyle{abbrv}

	\appendix

\section{Discussion}
\label{sec:discussion}

As shown above, the popular conjectures are not fine-grained enough for our purposes and our only viable option is to start from assumptions about the hardness of shaving logs for some problem.
The approach taken in this paper and in \cite{AHVW16} is to start with variants of SAT.
Another option would have been to conjecture that $3$\nobreakdash-SUM cannot be solved in $O(n^2/\log^{2+\eps}n)$ time, but SAT has several advantages.
First, SAT is deeply connected to fundamental topics in complexity theory, which allows us to borrow barriers that complexity theorists have faced for decades.
Moreover, there is a vast number of combinatorial problems that we can reduce SAT to, whereas $3$\nobreakdash-SUM seems more useful in geometric contexts, e.g. $3$-SUM-hardness for LCS and Frechet might be impossible \cite{BuchinBMM14}.
Thus, for the task of proving barriers for shaving logs, our approach seems as good as any.

Conditional lower bounds can even lead to better algorithms, by suggesting regimes of possible improvements. Phrased this way, our results leave the open problem of finding a $\Oh(n^2 / (\log n)^{2 + \eps})$ time algorithms for LCS, and perhaps more interestingly, shaving many more logs for the related-but-different Edit-Distance problem. The longstanding upper bound for Edit-Distance is $O(n^2/\log^2n)$ \cite{MP80} and our approach does not give barriers higher than $\Omega(n^2/\log^{20}n)$.

Finally, regardless of the consequences of our reductions, we think that the statements themselves are intrinsically interesting as they reveal a surprisingly close connection between Formula-SAT (a problem typically studied by complexity theorists) and combinatorial problems that are typically studied by stringologists, computational biologists, and computational geometers, which are \emph{a priori} completely different creatures. The runtime of the standard algorithm for SAT can be recovered almost exactly by encoding the formula into an LCS, Fr\'echet, or Pattern Matching instance and running the standard dynamic programming algorithms!

		\section{From Formula-SAT to Formula-Pair}
	\label{sec:simple}

In this section we show a chain of simple reductions starting from variants of Formula-SAT, which have $2^n$ time complexity, and ending at $n^2$ time variants of a problem we call \textsc{Formula-Pair}.

A formula $F$ of size $s$ over $n$ variables $x_1,\ldots,x_{n}$ is in the class $\mathcal{F}_1$ iff it has the following properties.
The gates in the first layer (nodes in the tree whose children are all leaves) compute arbitrary functions $C:\{0,1\}^{n/2} \to \{0,1\}$, as long as $C$ can be computed in $2^{o(n)}$ time and all children of a gate are marked with variables in $\{x_1,\ldots,x_{n/2}\}$ or with variables in $\{x_{n/2+1},\ldots,x_n\}$ but not with both.
W.l.o.g.\ we can assume that the inputs are only connected to nodes in the first layer.
The gates in the other layers compute deMorgan gates, i.e., OR and AND gates.
The size of $F$ is considered to be the \emph{number of gates in the first layer}. 
Since $F$ is a formula and thus has fanout 1, our size measure is up to constant factors equal to the total number of all gates except the inputs.
Note that the complexity of the functions in the first layer and their number of incoming wires, i.e.\ the number of leaves in the tree, do not count towards the size of $F$.

\defproblem{$\mathcal{F}_1$-Formula-SAT}
{Formula $F = F(x_1,\ldots,x_n)$ of size $s$ with $n$ inputs from the class $\mathcal{F}_1$}
{Exist $x_1,\ldots,x_n \in \{0,1\}$ such that $F(x_1,\ldots,x_n) = \true$?}
{$\Oh(2^{n} (s / n + 1))$, even restricted to $s \le n^{1+o(1)}$}

Note that many techniques developed for deMorgan formulas (and generalizations) are not applicable to $\mathcal{F}_1$-Formula-SAT, since the first layer is so general. In particular, techniques based on collapses by random restrictions~\cite{CKS14,CKKSZ15,Tal15} do not seem to work, as the first layer can be resistant to such collapses.

A simple algorithm achieves $O(2^n \cdot s)$ runtime: Preprocess $F$ to create a table that allows one to quickly lookup the value of each one of the first-layer gates on a given input. Constructing the table takes time $s \cdot 2^{n/2} \cdot 2^{o(n)}$, and after we have it, a brute-force SAT algorithm takes $O(2^n \cdot s)$ time.

However, one can improve upon this simple algorithm and obtain time $\Oh(2^{n} s / n)$. This is the best time complexity we are aware of, up to log-factors, unless we restrict the formula size to $s \le n^{2-\Omega(1)}$. Our tightest barriers for shaving logs for LCS can be based on the assumption that $\mathcal{F}_1$-Formula-SAT cannot be solved in time $\Oh(2^n /n^{1+\eps})$, for any $\eps>0$, even when $s=n^{2+o(1)}$. Such upper bounds seem out of reach of current techniques.

\begin{thm}
  $\mathcal{F}_1$-Formula-SAT can be solved in time $\Oh(2^{n} (s / n + 1))$.
\end{thm}
\begin{proof}
  This follows by combining Lemma~\ref{lem:redfsatsuccpair} and Theorem~\ref{lem:algosuccpair} below, noting that due to the change in the machine model we lose one log-factor. It is also easy to directly design an algorithm with the claimed running time by following the ideas in Theorem~\ref{lem:algosuccpair}, noting that reading a packed word now takes time $\Oh(n)$ instead of $\Oh(1)$, and all accesses in the (exponential size) precomputed table now take time $\Oh(n)$. 
\end{proof}

Next, we move from $2^n$ to $n^2$ with a simple split-and-list reduction (similar to \cite{Wil05}) to the following ``pair finding" problem.

\defproblem{\textsc{Formula-Pair}}
{A deMorgan Formula $F = F(x_1,\ldots,x_m,y_1,\ldots,y_m)$ of size $2m$ where each input is used exactly once, and $A,B \subseteq \{0,1\}^m$ of size $n$}
{Exist $a \in A, b \in B$ such that $F(a,b) = F(a_1,\ldots,a_m,b_1,\ldots,b_m) = \true$?}
{$\Oh(n^2 m / \log^2 n + n^2 / \log n)$ (Theorem~\ref{lem:algosuccpair}), even restricted to $m \le (\log n)^{1+o(1)}$}

We remark that the assumption that $F$ reads any input exactly once is w.l.o.g., since the sets $A,B$ allow us to ``copy'' an input $x_i$ to $x_j$ by simply ensuring that $a_i = a_j$ for all $a \in A$.

\begin{lem} \label{lem:redfsatsuccpair}
An instance of $\mathcal{F}_1$-Formula-SAT on a formula of size $s$ over $n$ inputs can be reduced to an instance of \textsc{Formula-Pair} on two sets of size $O(2^{n/2})$ and a formula of size $m=O(s)$, in linear time.
\end{lem}

\begin{proof}
Let $C_1,\ldots,C_m$ be the gates of the first layer of $F$ that compute functions of the first half of the variables, and let $C_1',\ldots,C_m'$ be the rest of the gates of the first layer, for some $k \leq s$.
For each partial assignment to the first $n/2$ variables of the formula we compute a bit-string $a \in \{0,1\}^m$ such that $a_i$ is $1$ iff $C_i$ outputs $1$ on the corresponding partial assignment.
Similarly, we compute a bit-string $b \in \{0,1\}^m$ for each partial assignment to the second half of the variables.
We define our deMorgan formula $F'$ to be equivalent to the layers of $F$ that are above the first one.
Now, to conclude the reduction, we observe that for any two partial assignments $\alpha,\beta$ to $F$, the corresponding bit strings $a,b$ contain the values of the first layer of $F$ on the assignment $(\alpha\beta)$ and so the value of $F'$ on $(ab)$ is exactly the value of $F$ on $(\alpha\beta)$.
Thus, $F$ is satisfiable iff there is a pair $a \in A, b \in B$ that satisfies $F'$. Since $F$ has fanout 1, every input of $F'$ is used exactly once.
\end{proof}

A trivial algorithm for \textsc{Formula-Pair} takes $O(n^2 \cdot m)$ time, and in Section~\ref{sec:algosuccpair} we show how to shave two log factors to $O(n^2 m/ \log^2{n})$, for $m = \Omega(\log n)$. Via the above reduction, this yields an $\Oh(2^{n} s / n)$ algorithm for $\mathcal{F}_1$-Formula-SAT for $s = \Omega(n)$, due to the loss of a $\log(n)$ factor in the jump from $2^n$ to $n$ and the machine model.

In the other sections we reduce \textsc{Formula-Pair} to LCS and show that any improvement by more than another log-factor on its running time directly yields improved runtimes for all problems above, including $\mathcal{F}_1$-Formula-SAT.
While we chose to present the barriers for shaving logs for LCS in terms of barriers for solving SAT faster, it is only safer to conjecture that the possibly harder \textsc{Formula-Pair} problem cannot be solved faster.

Next, we consider different variants that we will be able to reduce to the Fr\'echet distance.
A formula $F$ of size $s$ over $n$ variables $x_1,\ldots,x_{n}$ is in the class $\mathcal{F}_2$ iff it has the following properties.
The gates in the first layer (nodes in the tree whose children are all leaves) are as in $\mathcal{F}_1$. The gates in the second layer compute threshold functions $\tau:\{0,1\}^n \to \{0,1\}$ of their inputs, i.e. $\tau(x_1,\ldots,x_n) = \true$ iff $\sum_{i=1}^n c_i x_i \leq T$ where $c_1,\ldots,c_n,T \in \{-M,-M+1\ldots,M\}$.
W.l.o.g.\ we can assume that the inputs are only connected to nodes in the first layer, and first layer nodes are only connected to second layer nodes.
The gates in the other layers compute deMorgan gates, i.e., OR and AND gates.
The size of $F$ is considered to be the \emph{number of gates in the first plus second layer}.
Note that the complexity of the functions in the first layer and their number of incoming wires, i.e. the number of leaves in the tree, do not count towards the size of $F$.

\defproblem{$\mathcal{F}_2$-Formula-SAT}
{Formula $F = F(x_1,\ldots,x_n)$ of size $s$ with $n$ inputs from the class $\mathcal{F}_2$ with coordinates bounded by $M \le 2^{\Oh(n)}$}
{Exist $x_1,\ldots,x_n \in \{0,1\}$ such that $F(x_1,\ldots,x_n) = \true$?}
{$\Oh(2^{n} s \cdot n^{o(1)})$, even restricted to $s \le n^{1+o(1)}$}

The $\Oh(2^{n} s \cdot n^{o(1)})$ algorithm follows from Lemma~\ref{lem:redfsatineqsuccpair} and the algorithm for \textsc{Ineq-Formula-Pair} below, again noting that we lose one log-factor due to the change in the machine model.
Our tight barriers for shaving logs for Fr\'echet will be based on the assumption that $\mathcal{F}_2$-Formula-SAT cannot be solved in $2^n /n^{\eps}$, for some $\eps>0$, even when $s = n^{1+o(1)}$.
Since $\mathcal{F}_2$ is even more expressive than $\mathcal{F}_1$, such upper bounds seem out of reach.

\textit{Remark:} Any formula in $\mathcal{F}_2$ (of size $s$ over $n$ variables) can be transformed into a formula in $\mathcal{F}_1$ over the same variables of size $\textup{poly}(s)$, since arithmetic has efficient circuits. However, since we care about polynomial factors in $s$ in this paper, the difference between $\mathcal{F}_1$ and $\mathcal{F}_2$ is non-negligible for our purposes. In fact, the fastest running times that we know for $\mathcal{F}_1$-Formula-SAT and $\mathcal{F}_2$-Formula-SAT differ by a factor $n^{1 + o(1)}$.

We will work with an intermediate ``pair finding" problem:

\defproblem{Ineq-Formula-Pair}
{Formula $\F = \F(x_1,\ldots,x_m)$ where each input is used exactly once, and $A,B \subseteq \{-M,\ldots,M\}^m$ of size $n$, where $M \le n^{\Oh(1)}$}
{Exist $a \in A, b \in B$ such that $\F([a_1\le b_1],\ldots,[a_m \le b_m]) = \true$?}
{$\Oh(n^2 m\frac{\log \log n}{\log n})$, even restricted to $m \le (\log n)^{1+o(1)}$}

Similarly to the reductions above, we can prove the following. Similar reductions have been used in~\cite{Wi14,ILPS14}.

\begin{lem} \label{lem:redfsatineqsuccpair}
An instance of $\mathcal{F}_2$-Formula-SAT on a formula of size $s$ over $n$ inputs can be reduced to an instance of \textsc{Ineq-Formula-Pair} on two sets of size $O(2^{n/2})$ and a formula of size $m=O(s)$, in linear time.
\end{lem}

  The claimed running time $\Oh(n^2 m\frac{\log \log n}{\log n})$ follows by combining our reduction from \textsc{Ineq-Formula-Pair} to the \Fr distance with the fastest known algorithm for deciding the \Fr distance~\cite{AgarwalBAKS13}. We think that one can also obtain this running time more directly by ``Four Russians'' tricks, as have been used for LCS for large alphabet size. The further tricks that shaved a second log-factor off the running time of \textsc{Formula-Pair} do not seem to work for \textsc{Ineq-Formula-Pair}.
%

\subsection{Algorithm for \textsc{Formula-Pair}}
\label{sec:algosuccpair}

\begin{thm} \label{lem:algosuccpair}
  \textsc{Formula-Pair} can be solved in time $\Oh(n^2 m / \log^2 n + n^2 / \log n)$.
\end{thm}

The main trick used to prove the above theorem is the following decomposition of formulas.

\begin{lem} \label{lem:formulapartitioning}
  Given a deMorgan formula $F$ of size $m$ and a number $L \ge 2$, in polynomial time we can compute a decoposition $\mathcal{D} = \{F_1,\ldots,F_k\}$ such that
  \begin{enumerate}
    \item each $F_i$ is a subformula of $F$, i.e., a set of gates of $F$ forming a subtree,
    \item for each wire $e$ of $F$, exactly one $F_i$ contains both endpoints of $e$,
    \item each $F_i$ contains less that $3L$ gates, and
    \item $|\mathcal{D}| \le 4m/L + 1$.
  \end{enumerate}
  Furthermore, call a gate in $F_i$ \emph{special} if it is the root of another subformula $F_j$, $j \ne i$. In our decomposition each $F_i$ has at most 2 special gates.
\end{lem}

Similar decompositions where known before, see, e.g.,~\cite{dantsin2013exponential}, however, we are not aware of a decomposition bounding the number of special gates by a constant.

\begin{proof}
  Note that we can assume $F$ to have indegrees bounded by 2.
  Initialize $\mathcal{D} = \emptyset$.
  We assign to each gate $g$ of $F$ a weight $w(g)$ in $\{1,L\}$; initially all weights are 1, and inner nodes will always have weight 1. Now repeatedly perform the following procedure. Start at the root of $F$ and repeatedly go to the child whose subtree has larger total weight, until reaching a gate $g$ whose subtree has total weight in $[\tfrac 54 L, 3 L)$. 
  Add the subformula computed by $g$ to $\mathcal{D}$. Finally, remove all gates below $g$ from $F$ and set the weight of $g$ to $L$. Repeat this procedure until the total weight of $F$ is less than $\tfrac 54 L$. Add the remaining formula $F$ to $\mathcal{D}$. This defines a decomposition $\mathcal{D}$.
   
  Note that we always find a gate whose subtree has total weight in $[\tfrac 54 L, 3 L)$. Indeed, if the current gate $g$ has total subtree weight $w \ge 3L$, then $g$ is an inner node and thus has weight 1, and $g$'s children together have total subtree weight at least $w-1$. Since we go to the heavier child, we obtain total subtree weight at least $(w-1)/2 \ge (3L-1)/2 \ge \tfrac 54 L$, where we used $L \ge 2$. Hence, we will land in the interval $[\tfrac 54 L, 3 L)$.
  
  Also note that $|{\mathcal{D}}| \le 4m/L + 1$, since in each call of the procedure we remove weight at least $\tfrac 54 L$ (by deleting the gates below $g$) and then increase the weight of the chosen gate $g$ by at most $L$, so we lose weight at least $L/4$ in every call (except possibly in the very last step).
  
  Let ${\mathcal{D}} = \{F_1,\ldots,F_k\}$. Each subformula $F_i$ of $F$ has total weight less than $3L$, and thus it consists of less than $3L$ gates. Moreover, the gates $g$ in $F_i$ with weight $L$ exactly correspond to the special gates of $F_i$. Since each such gate $g$ has weight $L$, there are less than $3L / L$ special gates, and thus each $F_i$ has at most 2 special gates. 
\end{proof}

\begin{proof}[Proof of Theorem~\ref{lem:algosuccpair}]
  To solve \textsc{Formula-Pair} in time $\Oh(n^2 m / \log^2 n + n^2 / \log n)$, we use the ``Four Russians" trick as well as packing $\log(n)$ bits in a word.
  Let ${\mathcal{D}} = \{F_1,\ldots,F_k\}$ be a decomposition of $F$ as in Lemma~\ref{lem:formulapartitioning}, where $L := \eps \log n$, for a sufficiently small constant $\eps > 0$. Note that each subfomula $F_i$ has size $O(\eps \log n)$ and thus contains $O(\eps \log n)$ input gates of $F$.
  
  For each subformula $F_i$ and each assignment $\beta$ to the $\Oh(\eps \log n)$ $b$-variables appearing in $F_i$, let $F_i^\beta$ be the simplified subformula after fixing the $b$-variables. Compute the value of $F_i^\beta$ on all vectors in $A$ and all possible assignments to the special gates in $F_i$. We store these values in memory as follows. For each $1 \le \ell \le n$ divisible by $L$ and all length-$L$ bitstrings $x,x'$, store the values of $F_i^\beta$ on the $\ell+j$-th vector in $A$, with the input of the special gates set to $x_j,x'_j$ (recall that there are at most 2 special gates). Store these $L$ bits, for $0 \le j < L$, in one word. 
  We thus stored for each of $O(m / \log n)$ subformulas, each of $n^{\Oh(\eps)}$ assignments to the $b$-variables, each of $\Oh(n/L)$ bundles of vectors in $A$ (indexed by $\ell$), and each of $n^{\Oh(\eps)}$ assignments to the special gates, one word in memory. Computing these values takes time $m n^{1+\Oh(\eps)}$ and is thus negligible.
  
  Now we can evaluate $F$ on all pairs $a \in A$, $b \in B$ as follows. Iterate over all $b \in B$, and over all bundles of $L$ consecutive vectors $A' \subseteq A$ (indexed by $\ell$, divisible by $L$). Iterate over the subformulas $F_i$ in a topological order (from bottom to top). For each subformula $F_i$, if $F_i$ has no special gates then access $F_i^\beta$, the simplification of $F_i$ under $b$, and determine by one table lookup the output of $F_i^\beta$ on all vectors $a \in A'$. If $F_i$ does have special gates, then it depends on the output of some subformulas $F_j$, for which we have already computed the output on $b$ and any $a \in A'$. Plugging in these outputs as the vectors $x,x'$ above, we can again determine by one table lookup the output of $F_i^\beta$ on all vectors $a \in A'$. Repeating this for all subformulas $F_i$ eventually yields the output of $F$ on $b$ and any $a \in A'$. 
  
  Note that for each vector $b$ and bundle $A'$ this procedure takes time $\Oh(|{\mathcal{D}}|) = \Oh(m / \log n + 1)$. Since there are $n$ vectors $b$ and $\Oh(n/\log n)$ bundles $A'$, the total running time is $\Oh(n^2 m / \log^2 n + n^2 / \log n)$. In the end, we check whether the output on any pair $a \in A, b \in B$ is 1 to decide the given \textsc{Formula-Pair} instance.
\end{proof}

	\section{Missing Details in the LCS Proof}
	\label{sec:lcsproofs}
	This section contains the missing details for completing the reduction from \textsc{Formula-Pair} to LCS that we presented in Section~\ref{sec:technical}.

We will make use of the fact that for any strings $x,y$, symbol $c$, and $k \ge 1$ we have $\LCS(1^k x, 1^k y) = \LCS(x 1^k, y 1^k) = k + \LCS(x,y)$, which allows to greedily match prefixes and suffixes (see, e.g.,~\cite[Fact 7.1.(1)]{BK15}).

\begin{proof}[Proof of Lemma~\ref{lem:OR}]
  Clearly, we have $L \ge \LCS(0^\beta 1^\beta,0^\beta1^\beta) + \LCS(x_1',y_1')+ \LCS(2^\beta3^\beta,2^\beta3^\beta) = 4\beta + L'_1$ and $L \ge \LCS(2^\beta3^\beta,2^\beta3^\beta) + \LCS(x_2',y_2') + \LCS(0^\beta1^\beta,0^\beta1^\beta) = 4\beta + L'_2$. For the other direction, note that no LCS of $x,y$ can match symbols both in the $0^\beta1^\beta$-prefix of $x$ and in the $2^\beta3^\beta$-prefix of $y$, since these prefixes use disjoint alphabets and thus any such matchings would cross. Hence, 
  \begin{align} \label{eq:lcsoreqone}
    L \le \max\big\{ \LCS( x_1' 2^\beta3^\beta x_2' 0^\beta1^\beta,\, y), \LCS(x,\, y_2' 0^\beta1^\beta y_1' 2^\beta3^\beta) \big\}. 
  \end{align}
  Consider the first term, $\LCS( \tilde x, y)$ for $\tilde x = x_1' 2^\beta3^\beta x_2' 0^\beta1^\beta$. By the same argument as above, no LCS of $\tilde x, y$ can match symbols both in the $0^\beta 1^\beta$-suffix of $\tilde x$ and in the $2^\beta3^\beta$-suffix of $y$. Thus,
  \begin{align} \label{eq:lcsoreqtwo}
    \LCS( \tilde x,\, y) = \max\big\{ \LCS( x_1' 2^\beta3^\beta x_2',\, y), \LCS( \tilde x,\, 2^\beta3^\beta y_2' 0^\beta1^\beta y_1') \big\}. 
  \end{align}
  Note that $\LCS( x_1' 2^\beta3^\beta x_2',\, y) \le \LCS(2^\beta3^\beta,\, y) + \LCS( x_1' x_2',\, y) = 2\beta + \LCS( x_1' x_2',\, y)$. We bound $\LCS( x_1' x_2',\, y)$ by summing up the LCS length of all pairs of a part $x' \in \{x_1', x_2'\}$ in the first string and a part $y' \in \{2^\beta, 3^\beta, y_2', 0^\beta, 1^\beta, y_1', 2^\beta, 3^\beta\}$ in $y$, noting that we can either match $x_1'$ and $y_1'$ or $x_2'$ and $y_2'$, so that the two pairs $(x_1',y_1'),(x_2',y_2')$ together contribute at most $\max\{L'_1,L'_2\}$. This yields 
  $$ \LCS( x_1' x_2',\, y) \le \max\{L'_1,L'_2\} + 2 \cdot \beta/8 + 12 \cdot \beta/48 \le \beta/2 + \max\{L'_1,L'_2\}. $$
  Together we obtain, as desired, 
  \begin{align} \label{eq:lcsoreqthree}
    \LCS( x_1' 2^\beta3^\beta x_2',\, y) \le \tfrac 52 \beta + \max\{L'_1,L'_2\}.
  \end{align}
  
  For the other term $\LCS(\tilde x,\, \tilde y) = \LCS( \tilde x,\, 2^\beta3^\beta y_2' 0^\beta1^\beta y_1')$ of \eq{lcsoreqtwo}, let $z$ be an LCS and consider the number $k$ of symbols in the $2^\beta3^\beta$-prefix of $\tilde y$ matched by $z$. Note that we can bound $\LCS(\tilde x,\, \tilde y) \le k + \LCS(\tilde x, y_1' 0^\beta1^\beta y_2') \le k + \LCS(x,\, y_1' 0^\beta1^\beta y_2')$. By symmetry, we can bound $\LCS(x,\, y_1' 0^\beta1^\beta y_2')$ in the same way as \eq{lcsoreqthree} by $\tfrac 52 \beta + \max\{L'_1,L'_2\}$. Thus, for $k \le \tfrac 32 \beta$ we obtain $\LCS(\tilde x,\, \tilde y) \le 4\beta + \max\{L'_1,L'_2\}$, as desired. 
  
  In the remaining case $k > \tfrac 32 \beta$, we argue that we can split $\LCS(\tilde x,\, \tilde y) = \LCS(x_1' 2^\beta 3^\beta,\, 2^\beta3^\beta) + \LCS(x_2' 0^\beta1^\beta,\, y_2' 0^\beta1^\beta y_2')$. Note that at least $k-\beta > \beta/2$ symbols 2 of the $2^\beta3^\beta$-prefix of $\tilde y$ are matched. Since $x_1'$ and $x_2'$ contain at most $\beta/48$ symbols 2 by assumption, at least one symbol of the $2^\beta$-block in $\tilde y$ is matched to the $2^\beta$-block in $\tilde x$. Similarly, it follows that at least one symbol of the $3^\beta$-block in $\tilde y$ is matched to the $3^\beta$-block in $\tilde x$. Thus, the LCS $z$ splits between $2^\beta$ and $3^\beta$, i.e., we have 
  $$ \LCS(\tilde x,\, \tilde y) = \LCS(x_1' 2^\beta,\, 2^\beta) + \LCS(3^\beta x_2' 0^\beta1^\beta,\, 3^\beta y_2' 0^\beta 1^\beta y_1'). $$
  Using greedy prefix matching, we can remove the prefix $3^\beta$ of the second term to get
  \begin{align*}
    \LCS(\tilde x,\, \tilde y) &= \LCS(x_1' 2^\beta,\, 2^\beta) + \LCS(3^\beta,\, 3^\beta) + \LCS(x_2' 0^\beta1^\beta,\, y_2' 0^\beta 1^\beta y_1')  \\
    &= \LCS(x_1' 2^\beta3^\beta,\, 2^\beta3^\beta) + \LCS(x_2' 0^\beta1^\beta, \,y_2' 0^\beta 1^\beta y_1'). 
  \end{align*}
  Hence, there exists an LCS of $\tilde x, \tilde y$ matching only symbols in the prefixes $x_1' 2^\beta 3^\beta$ and $2^\beta 3^\beta$ and symbols in the suffixes $x_2' 0^\beta1^\beta$ and $y_2' 0^\beta1^\beta y_1'$. By symmetry, we can also split off the suffixes $0^\beta1^\beta$ and $0^\beta1^\beta y_2'$. It follows that
  $$ \LCS(\tilde x,\, \tilde y) = \LCS(x_1' 2^\beta 3^\beta,\, 2^\beta3^\beta) + \LCS(x_2',\, y_2') + \LCS(0^\beta1^\beta,\, 0^\beta1^\beta y_2') = 4\beta + L'_2, $$
  as desired. This finishes the analysis of the first term in \eq{lcsoreqone}. The second term gives the symmetric bound $4\beta + L'_1$, so in total we obtain the desired bound $L \le 4\beta + \max\{L'_1,L'_2\}$.
\end{proof}

\begin{proof}[Proof of Lemma~\ref{lem:length}]
In the base case we construct strings of length 1, so that $n(g), \rho(g) =1$. 

For an inner gate $g$, we recursively construct strings of length $n_1,n_2$ and LCS bounds $\rho_1,\rho_2$ for the children on subformulas $\F_1,\F_2$.
If $g$ is an AND gate we have 
\begin{align*}
  n(g) &= n_1+n_2+2\beta = n_1+n_2+2\lceil (n_1+n_2)/\tau^2 \rceil \le (n_1 + n_2)(1 + 2/\tau^2) + 2,  \\
  \rho(g) &= \rho_1 + \rho_2 + 2\beta \le \rho_1 + \rho_2 + 2(n_1+n_2)/\tau^2 + 2.
\end{align*} 
If $g$ is an OR gate we pad the strings to the same LCS length, thus increasing $n(g)$ by $|\rho_1 - \rho_2|$,
\begin{align*}
  n(g) &= n_1 + n_2 + |\rho_1 - \rho_2| + 6\beta \le (n_1 + n_2)(1 + 6/\tau^2) + 6 + \rho_1 + \rho_2,  \\
  \rho(g) &= 4\beta + \max\{\rho_1, \rho_2\} \le \rho_1 + \rho_2 + 4(n_1 + n_2)/\tau^2 + 4.
\end{align*} 
In both cases, we have
\begin{align*}
  n(g) &\le (n_1 + n_2)(1 + 6/\tau^2) + 6 + \rho_1 + \rho_2,  \\
  \rho(g) &\le \rho_1 + \rho_2 + 4(n_1 + n_2)/\tau^2 + 4.
\end{align*} 
Plugging in the inductive hypothesis yields, since $\F_1,\F_2$ have depth at most $\depth(\F_g)-1$,
\begin{align*}
  n(g) &\le 6\tau (|\F_1| + |\F_2|) (1+7/\tau)^{\depth(\F_g)-1} (1 + 6/\tau^2) + 6 + 6 (|\F_1|+|\F_2|) (1+7/\tau)^{\depth(\F_g)-1}.
\end{align*} 
We use the fact $|\F_g| = |\F_1| + |\F_2| + 1$ to cancel the additive term $+6$ at the cost of increasing the first $|\F_1| + |\F_2|$ to $|\F_g|$. We obtain 
\begin{align*}
  n(g) &\le 6\tau |\F_g|  (1+7/\tau)^{\depth(\F_g)-1} (1 + 6/\tau^2)  + 6 |\F_g| (1+7/\tau)^{\depth(\F_g)-1}  \\
  &= \frac{1 + 6/\tau^2 + 1/\tau}{1+7/\tau} \cdot 6 \tau |\F_g|  (1+7/\tau)^{\depth(\F_g)}.
\end{align*} 
Since $\tau = (\log \sigma)^{1/4} \ge 1$, as $\sigma$ is sufficiently large, we have $1 + 6/\tau^2 + 1/\tau \le 1+7/\tau$, which yields the desired bound.
For $\rho(g)$ we similarly obtain
\begin{align*}
  \rho(g) &\le 6(|\F_1| + |\F_2|) (1+7/\tau)^{\depth(\F_g)-1} + 4\cdot 6\tau(|\F_1| + |\F_2|) (1+7/\tau)^{\depth(\F_g)-1}/\tau^2 + 4
\end{align*}
Again we cancel the additive $+4$ by increasing $|\F_1| + |\F_2|$ to $|\F_g|$. We obtain
\begin{align*}
  \rho(g) &\le 6 |\F_g| (1+7/\tau)^{\depth(\F_g)-1} + 4\cdot 6\tau |\F_g| (1+7/\tau)^{\depth(\F_g)-1}/\tau^2  \\
  &= \frac{1 + 4/\tau}{1+7/\tau} \cdot 6 |\F_g| (1+7/\tau)^{\depth(\F_g)},
\end{align*}
which yields the desired bound and finishes the proof.
\end{proof}

\subsection{The Outer OR - Finalizing the Proof} 
\label{sec:lcsalignmentgadgets}

We now finish the reduction from \SuccPair to LCS using alignment gadgets~\cite{ABV15,BK15}. The following result is implicit in~\cite[Lemmas 3 and 4]{ABV15}, and a similar (independently found) result is contained in~\cite{BK15} under the name of ``alignment gadget''.

\begin{lem} \label{lem:alignment}
  Let $x_1,\ldots,x_n$ and $y_1,\ldots,y_n$ be strings of length $\lambda$. Let $\rho \ge \LCS(x_i,y_j)$ for all $i,j$. 
  Then one can construct strings $x,y$ of length $\Oh( n \lambda)$ and an integer $\rho'$ such that (1) $\LCS(x,y) \le \rho'$ and (2) $\LCS(x,y) = \rho'$ if and only if there exist $i,j$ with $\LCS(x_i,y_j) = \rho$. If the strings $x_i,y_j$ use alphabet $\Sigma$, then $x,y$ use alphabet size at most $|\Sigma| + 4$.
\end{lem}

For any vectors $a \in A, b \in B$, invoking \lemref{lcsgatesconstr} for the output gate of $\F$, we obtain strings $x(a),y(b)$ and a number $\rho$ such that (1) $\LCS(x(a),y(b)) \le \rho$ and (2) $\LCS(x(a),y(b)) = \rho$ if and only if $\F(a,b)$ evaluates to true. Tracing the proof of \lemref{lcsgatesconstr} one can see that $\rho$ does not depend on the choice of $a,b$, and similarly $x(a)$ does not depend on $b$ and $y(b)$ does not depend on $a$. Hence, we obtain $n$ strings $x(a), a \in A$ and $n$ strings $y(b), b \in B$ and a threshold $\rho$ on which we can invoke \lemref{alignment}. This yields strings $x,y$ and a threshold $\rho'$ such that (1) $\LCS(x,y) \le \rho'$ and (2) $\LCS(x,y) = \rho'$ if and only if there exist $a \in A, b \in B$ such that $\F(a,b)$ evaluates to true, so indeed from the LCS of the constructed strings $x,y$ one can solve the \SuccPair instance.

The strings $x,y$ use alphabet size $5 \sigma^2 + 4$. Using the length bound of \lemref{lcsgatesconstr}, $x,y$ have length $\Oh\big(n \tau |\F| (1+7/\tau)^{\depth(\F)}\big)$. Here, as in \lemref{lcsgatesconstr}, we may choose any sufficiently large $\sigma > 0$ and set $\tau = (\log \sigma)^{1/4}$. 

Since a priori the depth of $\F$ could be as large as $|\F|$, the factor $(1+7/\tau)^{\depth(\F)}$ in our length bound is not yet satisfactory.
Thus, as a preprocessing before the above construction, we decrease the depth of $\F$ using the following result of Bonet and Buss \cite{Spira71,BB94}.

\begin{thm}[\cite{BB94}] \label{thm:bonetbuss}
  Let $\F$ be a formula. Then for all $k \ge 2$ there is an equivalent formula $\F'$ with depth at most $(3 k \ln 2) \log |\F|$ and size $|\F'| \le |\F|^{1 + 1/(1 + \log(k-1))}$. This fomula can be constructed in time $|\F|^{\Oh(1)}$. 
\end{thm}

Given an instance $(\F,A,B)$ of \SuccPair, we first run Bonet-Buss with $k := \sqrt{\tau}$ to obtain a formula $\F'$. Then we use the reduction to LCS on $\F'$. This yields an equivalent LCS instance $(x,y)$ of length $\Oh\big(n \tau |\F'| (1+7/\tau)^{\depth(\F')}\big)$, as shown above. Plugging in the bounds on $|\F'|$ from Bonet-Buss, we obtain length 
$$ \Oh\Big(n |\F|^{1 + 1/(1 + \log(k-1))} (1+7/\tau)^{(3 k \ln 2) \log |\F|}\Big) = \Oh\Big(n |\F|^{1 + 1/(1 + \log(k-1)) + \Oh(k/\tau)}\Big), $$
where we used $1 + x \le \exp(x) = 2^{\Oh(x)}$. Since we set $k = \sqrt{\tau}$, this evaluates to $\Oh\big(n |\F|^{1 + \Oh(1/\log \tau)}\big)$. Since furthermore $\tau = (\log \sigma)^{1/4}$, we obtain length bound $\Oh\big(n |\F|^{1 + \Oh(1/\log \log \sigma)}\big)$. This finishes the proof of \thmref{lcspair} for any sufficiently large alphabet size $5\sigma^2 + 4$. Note that it suffices to prove \thmref{lcspair} for sufficiently large alphabet size, since for any constant $C \ge 2$ and any alphabet size $|\Sigma| \le C$ the statement follows from the reduction in~\cite{AHVW16}, which has length bound $\Oh(n |\F|^{\Oh(1)})$ (with an explicit constant $\Oh(1)$ in the exponent).

Now assume that LCS on strings of length $n$ over sufficiently large alphabet size $\sigma$ has an algorithm in time $T(n) = \Oh(n^2 / (\log n)^{2+C/\log \log \sigma + \eps})$, for some sufficiently large $C > 0$ and any (small) $\eps > 0$. Then from the above reduction we obtain an algorithm for \SuccPair on formulas of size $|\F| \le (\log n)^{1+o(1)}$ in time 
$$ T\Big(n (\log n)^{1 + \Oh(1/\log \log \sigma) + o(1)}\Big) = \Oh\Big( n^2 (\log n)^{2 + \Oh(1/\log \log \sigma) + o(1)} / (\log n)^{2 + C/\log \log \sigma + \eps}\Big), $$
which is bounded by $\Oh\big(n^2 / (\log n)^{\eps/2}\big)$.
This would break the \SuccPair barrier, so we obtain a conditional lower bound for LCS of $n^2 / (\log n)^{2 + \Oh(1/\log \log \sigma) + o(1)}$. This implies \corref{lcs}.

	\section{Simple Reduction to LCS using Large Alphabet}
	\label{sec:lcssimple}


In this section we present a simpler alternative to the reduction from \textsc{Formula-Pair} to LCS given in Sections~\ref{sec:technical} and \ref{sec:lcsproofs}, using however a much larger alphabet size. 

\begin{thm}
\label{thm:lcspairsimple}
\textsc{Formula-Pair} on formulas of size $s$ and lists of size $n$ can be reduced to an instance of LCS on two strings over alphabet of size $\Theta(s)$ of length $O(n \cdot s \cdot 2^{\Oh(\sqrt{\log s})})$, in linear time. 
\end{thm}

Similarly as before, we construct strings $x,y$ and a number $\rho$ such that $\LCS(x,y) \ge \rho$ holds if and only if the given \SuccPair instance $(\F,A,B)$ is satisfiable. 
Fix vectors $a,b \in \{0,1\}^m$ (where $2m$ is the number of inputs to $\F$ and thus $m = \Theta(s)$). For any gate $g$ of $\F$, we construct strings as follows.

\begin{lem} \label{lem:lcsgatesconstrsimple}
  We can inductively construct, for each gate $g$ of $\F$, strings $x(g)=x(g,a)$ and $y(g)=y(g,b)$ over alphabet size $|F_g|$ (where $\F_g$ is the subformula of $\F$ below $g$) and a number $\rho(g)$ such that for $L(g) := \LCS(x(g),y(g))$ we have $L(g) \le \rho(g)$, with equality if and only if gate $g$ evaluates to true on input $(a,b)$ to $\F$. Moreover, we have $|x(g)| = |y(g)| = n(g) \le (\depth(F_g)+1) \cdot |\F_g|$.
\end{lem}

\begin{proof}
Consider any gate $g$ of $\F$, and write for readability $x = x(g)$ and similarly define $y,n,L,\rho$.
If $g$ has fanin 2, write $g_1,g_2$ for the children of $g$. Moreover, let $x_1 = x(g_1)$ and similarly define $y_1,n_1,L_1,\rho_1$ and $x_2,y_2,n_2,L_2,\rho_2$. 

\paragraph*{Input Gate}
The base case is an input bit $a_i$ to $\F$ (input bits $b_j$ are symmetric).
Interpreting $a_i$ as a string of length~1 over alphabet $\{0,1\}$, note that $\LCS(a_i,1) = a_i$. Hence, the strings $x = a_i$ and $y = 1$, with $n=\rho=1$, trivially simulate the input bit $a_i$.

\paragraph*{AND Gates}
For a gate $g = (g_1 \wedge g_2)$ we first replace all symbols in $x_2,y_2$ by fresh symbols, so that $x_1 y_1$ and $x_2 y_2$ use disjoint alphabets. We then construct
\begin{alignat*}{6}
  x \; &:= \;\; &&x_1 \;&& x_2 \\
  y \; &:= \;\; &&y_1 \;&& y_2.
\end{alignat*}
Since these strings use disjoint alphabets, it is easy to see that $\LCS(x,y) = \LCS(x_1,y_1) + \LCS(x_2,y_2)$. Hence, inductively we obtain for $\rho := \rho_1 + \rho_2$ that $L = \LCS(x,y) \le \rho$, with equality if and only if both gates $g_1$ and $g_2$ evaluate to true. Thus, we correctly implemented an AND gate. Note that $n = n_1 + n_2$.

\paragraph*{OR Gates}
For a gate $g = (g_1 \vee g_2)$ we first replace all symbols in $x_2,y_2$ by fresh symbols, so that $x_1 y_1$ and $x_2 y_2$ use disjoint alphabets. 
Without loss of generality assume that $\rho_1 \ge \rho_2$ (the other case is symmetric). Let $\delta := \rho_1 - \rho_2$ and let $\$$ be a fresh symbol. We construct the strings
\begin{alignat*}{6}
  x \; &:= \;\; &&x_1 \;&& x_2 \; \$^\delta \\
  y \; &:= \;\; &&y_2 \; \$^\delta \;&& y_1.
\end{alignat*}
Since $x_1,y_1$ and $x_2 \$^\delta, y_2 \$^\delta$ use disjoint alphabets, we can either match symbols in the first pair or in the second. It follows that $\LCS(x,y) = \max\{\LCS(x_1,y_1), \LCS(x_2 \$^\delta, y_2 \$^\delta)\}$. Since $\$$ is a fresh symbol, we moreover have $\LCS(x_2 \$^\delta, y_2 \$^\delta) = \LCS(x_2,y_2) + \delta$. Together, we obtain $L = \LCS(x,y) \le \max\{\rho_1, \rho_2 + \delta\} = \rho_1$, with equality if and only if $g_1$ or $g_2$ evaluate to true. Thus, we correctly implemented an OR gate. Note that $n = n_1 + n_2 + |\rho_1 - \rho_2|$ and $\rho = \max\{\rho_1, \rho_2\}$.

\paragraph*{Analyzing the Length} 
Note that both for AND and OR gates, the inequality $\rho \le \rho_1 + \rho_2$ holds, with the base case $\rho=1$ at input gates. This yields $\rho(g) \le |F_g|$. 
Furthermore, for both AND and OR gates the inequality $n \le n_1 + n_2 + |\rho_1  -\rho_2|$ holds. Since $|\rho_1  -\rho_2| \le \rho_1 + \rho_2 \le |F_{g_1}| + |F_{g_2}| \le |F_g|$, we obtain $n \le n_1 + n_2 + |F_g|$. 
It now follows inductively that $n \le (\depth(F_g)+1) \cdot |F_g|$, since 
\begin{align*}
  n \le n_1 + n_2 + |F_g| &\le  (\depth(F_{g_1})+1) \cdot |F_{g_1}| + (\depth(F_{g_2})+1) \cdot |F_{g_2}| + |F_g| \\
&\le \depth(F_g) \cdot( |F_{g_1}| + |F_{g_2}| ) + |F_g| \le (\depth(F_g)+1) \cdot |F_g|.
\end{align*}
Also note that since each gate can introduce one new symbol the alphabet size is at most $|F_g|$. This proves the lemma.
\end{proof}

\paragraph*{Outer OR} 
We now finish the reduction from \SuccPair to LCS using alignment gadgets, see \lemref{lcsgatesconstrsimple}.
For any vectors $a \in A, b \in B$, invoking \lemref{lcsgatesconstrsimple} for the output gate of $\F$, we obtain strings $x(a),y(b)$ and a number $\rho$ such that $\LCS(x(a),y(b)) \le \rho$, with equality if and only if $\F(a,b)$ evaluates to true. Tracing the proof of \lemref{lcsgatesconstrsimple} one can see that $\rho$ does not depend on the choice of $a,b$, and similarly $x(a)$ does not depend on $b$ and $y(b)$ does not depend on $a$. Hence, we obtain $n$ strings $x(a), a \in A$ and $n$ strings $y(b), b \in B$ and a threshold $\rho$ on which we can invoke \lemref{alignment}. This yields strings $x,y$ and a threshold $\rho'$ such that $\LCS(x,y) \le \rho'$, with equality if and only if there exist $a \in A, b \in B$ such that $\F(a,b)$ evaluates to true, so indeed from the LCS of the constructed strings $x,y$ one can solve the \SuccPair instance. Note that so far the resulting strings have length $\Oh(n \cdot \depth(\F) \cdot |\F|)$.

\paragraph*{Finalizing the Proof} 
Since a priori the depth of $\F$ could be as large as $|\F|$, the factor $\depth(\F)$ in our length bound is not yet satisfactory.
Thus, as a preprocessing before the above construction, we use the depth reduction of Bonet and Buss, see Theorem~\ref{thm:bonetbuss}.
Given an instance $(\F,A,B)$ of \SuccPair, we first run Bonet-Buss with $k := 2^{\Theta(\sqrt{\log(|F|)})}$ to obtain a formula $\F'$. Then we use the reduction to LCS on $\F'$. This yields an equivalent LCS instance $(x,y)$ of length $\Oh(n \cdot (\depth(\F')+1) \cdot |\F'|)$, as shown above. Plugging in the bounds on $|\F'|$ from Bonet-Buss, we obtain length $O(n \cdot |\F| \cdot 2^{\Theta(\sqrt{\log(|F|)})}) = \Oh(n \cdot |\F|^{1+o(1)})$.
This finishes the proof of Theorem~\ref{thm:lcspairsimple}.

	\section{The Reduction to Pattern Matching}
	\label{sec:patternmatch}

In this section we reduce the \textsc{Formula-Pair} problem to regular expression pattern matching. By the reductions in Section~\ref{sec:simple}, this gives us barriers based on $\mathcal{F}_1$-Formula-SAT and proves Theorem~\ref{thm:patternmatch}.
Let us start with a formal definition of the problem.

\paragraph*{Regular Expression Pattern Matching}
Regular expressions over alphabet $\Sigma$ are inductively defined as follows. For any $c \in \Sigma$, the regular expression $c$ matches only itself, i.e., $L(c) = \{c\}$. The regular expression $r = r_1 \circ r_2$ matches all concatenations of strings matched by $r_1$ and $r_2$, i.e., $L(r) = \{ s_1 s_2 \mid s_1 \in L(r_1), s_2 \in L(r_2) \}$. We often abbreviate $r_1 \circ r_2$ by $r_1 r_2$. The regular expression $r = r_1 | r_2$ matches the union of all strings matched by $r_1$ and $r_2$, i.e., $L(r) = L(r_1) \cup L(r_2)$. Finally, the regular expression $r = r_1^*$ matches all sequences of strings matched by $r_1$, i.e., $L(r) = \{s_1 \ldots s_k \mid k \ge 0, s_1, \ldots, s_k \in L(r_1)\}$. The \emph{regular expression pattern matching problem} is, given a string $t$ (the text) and a regular expression $p$ (the pattern), to decide whether $p$ matches some substring of $t$.

\medskip 
We will prove the following theorem. Together with the reductions in Section~\ref{sec:simple}, this implies Theorem~\ref{thm:patternmatch}.

\begin{thm}
\textsc{Formula-Pair} on formulas of size $s$ and lists of size $n$ can be reduced to an instance of regular expression pattern matching with text length and pattern size $O(n \cdot s \log s)$, in linear time. The alphabet is $\Sigma = \{0,1\}$.
\end{thm}

In our construction, we will use the following encoding of numbers. For any integer $m \ge 0$, let $m_1 \ldots m_k$ be its representation in binary, where $k = \max\{1, \lfloor \log_2 m \rfloor + 1\}$. We define the string
$$\#(m) := 1 \, 1 \, 0 \, m_1 \, 0 \, m_2 \, 0 \ldots 0 \, m_{k-1} \, 0 \, m_k \, 0 \, 1 \, 1.$$
We will make use of the following simple observation: For any $x, y_1,\ldots,y_\ell \ge 0$ the string $\#(x)$ is a substring of the concatenation $\#(y_1) \ldots \#(y_k)$ if and only if we have $x = y_i$ for some $i$. (In particular, $\#(x)$ cannot appear as a suffix of $\#(y_1)$ followed by a prefix of $\#(y_2)$.)

For each gate $g$ in the formula $\F$ and vectors $a,b \in \{0,1\}^k$ we design a text $t(g,a)$ and a pattern $p(g,b)$ such that $p(g,b)$ matches $t(g,a)$ if and only if on input $a,b$ in the formula $\F$ gate $g$ evaluates to true. In our construction we ensure that for any gate $g$ of height $h$ the text $t(g,a)$ can be written as $\#(y_1) \ldots \#(y_k)$ for some $k\ge 1$ and $0 \le y_1,\ldots,y_k \le h+1$.

In the following, fix a gate $g$ of $\F$ with children $g_1$ and $g_2$. For simplicity, we write $t = t(g,a), p = p(g,b)$ and similarly $t_1,p_1,t_2,p_2$ for $g_1$ and $g_2$. We denote by $h$ the height of $g$, i.e., the length of the longest path from $g$ to any of its descendants. 

\paragraph*{Input Gates} The base case is an input bit $a_i$ to $\F$ (input bits $b_j$ are symmetric). Since $a_i$ is a number in $\{0,1\}$, we can set $t := \#(a_i)$ and $p := \#(1)$. Then the pattern matches the text if and only if $a_i = 1$. Note that the height is 0 and we indeed only used $\#(0)$ and $\#(1)$.

\paragraph*{AND Gates} Consider a gate $g = (g_1 \wedge g_2)$ of height $h \ge 1$ and let $h' := h+1$. We set 
\begin{align*}
  t \;\;:=& \;\; t_1 \; \#(h') \; t_2, \\
  p \;\;:=& \;\; p_1 \; \#(h') \; p_2. 
\end{align*}
Since $t_1$ and $t_2$ only contain $\#(y)$ with $y < h'$, the only way for $p$ to match $t$ is to match the occurences of $\#(h')$ in $p$ and $t$ in the natural way, so that $p_1$ must match $t_1$ and $p_2$ must match $t_2$. Thus, we correctly implemented an AND gate.

\paragraph*{OR Gates} Consider a gate $g = (g_1 \vee g_2)$ of height $h \ge 1$ and let $h' := h+1$. We set 
\begin{align*}
  t \;\;:=&\;\; t_1 \; \#(h') \; t_2, \\
  p \;\;:=&\;\; p_1 \; \#(h') \; (0|1)^* \;\;\big|\;\; (0|1)^* \; \#(h') \; p_2.  
\end{align*}
This time we can choose which of the two sub-expressions of $p$ we want to match. In any case, there is an occurence of $\#(h')$ in the pattern, and since $t_1$ and $t_2$ only contain $\#(y)$ with $y < h'$, the only way for the pattern to match $t$ is to match the occurences of $\#(h')$ in $p$ and $t$ in the natural way. Since $(0|1)^*$ matches any string, in particular $t_1$ and $t_2$, we obtain that $p_1$ matches $t_1$ or $p_2$ matches $t_2$, depending on which sub-expression of $p$ we choose. Thus, we correctly implemented an OR gate.

This finishes the construction of the texts $t(g,a)$ and patterns $p(g,a)$. 

\paragraph*{Outer OR} Invoking this construction for the root $r$ of $\F$ yields a text $t(a) := t(r,a)$ and a pattern $p(b) := p(r,b)$ such that $p(b)$ matches $t(a)$ if and only if on input $a,b$ the formula $\F$ evaluates to true. Let $H$ be two plus the height of $r$, and let $A = \{a^1,\ldots,a^n\}$ and $B = \{b^1,\ldots,b^n\}$. We construct the final text and pattern as
\begin{align*}
  t \;\; =& \;\; \#(H) \; t(a^1) \; \#(H) \quad \ldots \quad \#(H) \; t(a^n) \; \#(H), \\
  p \;\; =& \;\; \#(H) \; p(b^1) \; \#(H) \;\, \big| \, \ldots \, \big| \,\; \#(H) \; p(b^n) \; \#(H).
\end{align*}
To match the pattern $p$, we have to choose one of the sub-expressions $p^j := \#(H) \, p(b^j) \, \#(H)$. Since $\#(H)$ does not appear in any $t(a^i)$, in order to match $p^j$ to a substring of $t$ we must match the two occurences of $\#(H)$ in $p^j$ to two top-level occurences of $\#(H)$ in $t$, and thus we must match $p(b^j)$ to some $t(a^i)$. As this is possible if and only if $\F$ is satisfied by $a^i,b^j$, we have constructed an instance of regular expression pattern matching that is equivalent to the given instance of \textsc{Formula-Pair}. 

Note that the height of any gate is bounded by $O(s)$, where $s$ is the size of the formula $\F$, and thus any encoding $\#(y)$ appearing in our construction has length $O(\log s)$. The text length and pattern length both satisfy a recursion of the type
$$ L \le L_1 + L_2 + O(\log s), $$
which yields the bound $O(s \log s)$ on the length of $t(a)$ and the size of $p(b)$. In total, we obtain length and size $O(n \cdot s \log s)$. This proves Theorem~\ref{thm:patternmatch}.

	\section{The Reduction to Fr\'echet Distance}
	\label{sec:frechet}

In this section we reduce the \textsc{Ineq-Formula-Pair} problem to the Fr\'echet distance. By the reductions in Section~\ref{sec:simple}, this gives us barriers based on \textsc{$\mathcal{F}_2$-Formula-SAT} and proves Theorem~\ref{thm:frechet}.
Familiarity with the SETH-hardness proof of \cite{Bring14} will help to understand our proof.
Let us start with a formal definition of the problem.

\paragraph*{\Fr distance}
A \emph{curve} $P$ is a sequence $(p_1,\ldots,p_n)$ of points in the Euclidean space $\R^d$. We call $p_1,\ldots,p_n$ the \emph{vertices} of $P$. In the construction in this paper, we will always work in dimension $d=2$, and the coordinates of vertices will be rationals with small bitlength. 
For two curves $P=(p_1,\ldots,p_n)$ and $Q=(q_1,\ldots,q_m)$ and any monotone non-decreasing and onto functions $\phi \colon \{1,\ldots,n+m\} \to \{1,\ldots,n\}$ and $\psi \colon \{1,\ldots,n+m\} \to \{1,\ldots,m\}$ we call $\tau = (\phi,\psi)$ a \emph{traversal}. We say that at time step $t$ the traversal $\tau$ is at vertex $p_{\phi(t)}$ in $P$ and at vertex $q_{\psi(t)}$ in $Q$. Observe that $\tau$ describes one way of a man and its dog walking along $P$ and $Q$, respectively, from their starting vertices to their ending vertices, where in each time step man and dog may step to the next vertex on their curve or stay at their current vertex. We say that the traversal $\tau$ \emph{stays in distance~$\delta$} if at any time step $1 \le t \le n+m$ the current vertices $p_{\phi(t)}$ and $q_{\psi(t)}$ are within distance~$\delta$ of each other. Finally, the \emph{\Fr distance $\dfr(P,Q)$} is the minimal number $\delta$ such that there exists a traversal $\tau$ of $P,Q$ staying in distance $\delta$.

\medskip 
We will prove the following theorem. Together with the reductions in Section~\ref{sec:simple}, this implies Theorem~\ref{thm:frechet}.

\begin{thm}
\textsc{Ineq-Formula-Pair} on formulas of size $s$ and lists of size $n$ can be reduced to an instance of \Fr on two curves of length $O(n \cdot s)$, in linear time. 
\end{thm}

\subsection{Outline of the Reduction}

We say that curves $P,Q$ are \emph{$\delta$-placed} if 
\begin{itemize}
  \item the $x$-coordinates of all vertices of $P$ and $Q$ are in $[-\delta,\delta]$,
  \item the $y$-coordinates of all vertices of $P$ are in $[1-\delta^2,1+\delta^2]$, and
  \item the $y$-coordinates of all vertices of $Q$ are in $[-\delta^2,\delta^2]$.
\end{itemize}
This ensures that $P$ is contained in a small region around the point $(1,0)$, and $Q$ is close to $(0,0)$.
\begin{lem} \label{lem:gates}
For each gate $g$ in the formula $\F$ and vectors $a,b \in \{-M,\ldots,M\}^k$, and any $\delta \in (0,1)$, we design curves $P=P_\delta(g,a)$ and $Q=Q_\delta(g,b)$ such that
\begin{enumerate}[label=(P\arabic*)]
\item $\dfr(P,Q) \le 1$ if and only if on input $a,b$ in the formula $\F$ gate $g$ evaluates to true, and \label{P1}
\item $P,Q$ are $\delta$-placed. \label{P2}
\end{enumerate}
The curves $P$ and $Q$ have $\Oh(|\F|)$ vertices. Coordinates of the vertices of $P,Q$ use $\Oh(\log(1/\delta) + h + \log M)$ bits, where $h$ is the height of gate $g$.
\end{lem}
Property \ref{P1} makes sure that we correctly simulate formula $\F$ by the \Fr distance. We use property \ref{P2} to gain control over the \Fr distance of recursively defined subcurves.
Before proving the above lemma, in \secref{outeror} we first design the ``outer OR'' of the reduction from \textsc{Ineq-Formula-Pair}, which chooses a pair of vectors $a \in A, b \in B$. We then show how to implement the different types of gates in \secref{gates}, proving \lemref{gates}.

\subsection{Outer OR} \label{sec:outeror}

Let $r$ be the root of formula $\F$ and consider the curves $P_a := P_{1/16}(r,a)$ and $Q_b := Q_{1/16}(r,b)$ for vectors $a \in A, b \in B$ given by \lemref{gates}. We want to construct curves $P,Q$ such that $\dfr(P,Q) \le 1$ if and only if there exist $a \in A, b \in B$ with $\dfr(P_a,Q_b) \le 1$, which holds if and only if $\F(a,b)$ evaluates to true. 

We study this situation slightly more generally as follows.
Let $\delta \in (0,1/2]$ and let $P_1,Q_1,\ldots,P_k,Q_k$ be $\delta/8$-placed curves. Define the following auxiliary points:
\begin{align*}
  s_P &:= (-\delta/2,1-\delta^2), & s_Q &:= (-\delta/2,\delta^2), & s_Q^* &:= (-\delta/2,-\delta^2),  \\
  t_P &:= (\delta/2,1-\delta^2), & t_Q &:= (\delta/2,\delta^2), & t_Q^* &:= (\delta/2,-\delta^2), \\
  b_P &:= (-\delta/2, 1), & b_Q &:= (-\delta/2,0), \\
  e_P &:= (\delta/2, 1), & e_Q &:= (\delta/2,0).
\end{align*}
We define the curve $P = P_\delta^{OR}(P_1,\ldots,P_k)$ as 
$$ P := \bigcirc_{\ell=1}^k s_P \circ b_P \circ P_\ell \circ e_P \circ t_P, $$
i.e. we start with vertices in $s_P$ and $b_P$, then follow $P_1$, then add vertices at $e_P, t_P, s_P$, and $b_P$, then follow $P_2$ and so on. We define $Q = Q_\delta^{OR}(Q_1,\ldots,Q_k)$ by
$$ Q := s_Q \circ s_Q^* \circ \big(\bigcirc_{\ell=1}^k b_Q \circ Q_\ell \circ e_Q \big) \circ t_Q^* \circ t_Q. $$
\begin{lem}[Outer OR] \label{lem:outeror}
  For any $\delta \in (0,1/2]$ and any $\delta/8$-placed curves $P_1,Q_1,\ldots,P_k,Q_k$, the curves $P = P_\delta^{OR}(P_1,\ldots,P_k)$ and $Q = Q_\delta^{OR}(Q_1,\ldots,Q_k)$ constructed above are $\delta$-placed and satisfy $\dfr(P,Q) \le 1$ if and only if there are $1 \le i,j \le k$ with $\dfr(P_i,Q_j) \le 1$.
\end{lem}

Note that for $\delta := 1/2$ the curves $P_a = P_{1/16}(r,a)$ and $Q_b = Q_{1/16}(r,b)$ satisfy the requirements of the above construction, in particular they are $(\delta/8 = 1/16)$-placed, and thus we obtain curves $P,Q$ such that $\dfr(P,Q) \le 1$ if and only if there exist $a \in A, b \in B$ with $\dfr(P_a,Q_b) \le 1$. This yields a reduction from \textsc{Ineq-Formula-Pair} on $n$ vectors and formula $\F$ to the \Fr distance on curves of length $\Oh(n |\F|)$, as $P_a,Q_b$ have length $\Oh(|\F|)$ by \lemref{gates}. The hypothesis that for $|\F| = (\log n)^{1+o(1)}$ \textsc{Ineq-Formula-Pair} has no $n^2 / (\log n)^{\eps}$ algorithm now implies that (the decision variant of) the \Fr distance has no $n^2 / (\log n)^{2+\eps'}$ algorithm, as desired. Also note that the coordinates of the vertices of the resulting curves use $\Oh(\depth(\F) + \log M)$ bits. Since by Bonet and Buss' depth reduction (\thmref{bonetbuss}) we can assume that $\depth(\F) = \Oh_\eps(\log |\F|)$ and $|\F| = (\log n)^{1+o(1)}$, these coordinates fit into a memory cell for any $M = n^{\Oh(1)}$. Thus, the reduction not only works on the Real RAM,  but also on the Word RAM. 

\begin{proof}[Proof of \lemref{outeror}]
  From the definition of the auxiliary points and since $P_\ell,Q_\ell$ are $\delta/8$-placed for all $\ell$, it follows that $P,Q$ are $\delta$-placed. 
  
  If there are $i,j$ such that $\dfr(P_i,Q_j) \le 1$ then we find a traversal of $P,Q$ staying in distance 1 as follows. Denote by $s_P^{(\ell)}$ the $\ell$-th occurrence of $s_P$ on curve $P$, i.e. the copy of $s_P$ that comes before $P_\ell$ in $P$, and similarly define $t^{(\ell)}_P, b^{(\ell)}_P, e_P^{(\ell)}, b^{(\ell)}_Q, e_Q^{(\ell)}$. Note that $s_Q,s_Q^*,t_Q,t_Q^*$ appear only once in $Q$, so they do not get a superscript. Any traversal of $P,Q$ starts in $s_P^{(1)}$ and $s_Q$. We stay at $s_Q$ and walk along $P$ until we reach $s_P^{(i)}$, staying in distance 1 by claim \ref{U1} below. Then we stay in $s_P^{(i)}$ and walk along $Q$ until we reach $b_Q^{(j)}$, using \ref{U2}. Now we step to $b_P^{(i)}$ in $P$, using \ref{U4}. After a simultaneous step in $P,Q$ we are at the starting vertices of $P_i,Q_j$ and we proceed by optimally traversing them, staying in distance 1 since $\dfr(P_i,Q_j) \le 1$ by assumption. We then make a simultaneous step to $e_P^{(i)}$ and $e_Q^{(j)}$, using \ref{U4}. Now we make a single step in $P$ to $t_P^{(i)}$, using \ref{U3}. We stay in this vertex of $P$ and walk along $Q$ until its final vertex $t_Q$, using \ref{U3}. Finally, we stay in $t_Q$ and walk along $P$ until its final vertex $t_P^{(k)}$, using \ref{U1}. It remains to show the following.
  
  \begin{claim} We have the following upper bounds on vertex distances:
  \begin{enumerate}[label=(U\arabic*)]
    \item $s_Q$ and $t_Q$ are in distance 1 of any vertex of $P$, \label{U1}
    \item $s_P$ is in distance 1 of any vertex of $Q$ except for $t_Q^*$, \label{U2}
    \item $t_P$ is in distance 1 of any vertex of $Q$ except for $s_Q^*$, \label{U3}
    \item $b_P$ and $b_Q$ as well as $e_P$ and $e_Q$ are in distance 1. \label{U4}
  \end{enumerate}
  \end{claim}
  \begin{proof}
  For \ref{U1} consider $s_Q$, since $t_Q$ is symmetric. Observe that every vertex on $P$ has $y$-coordinate at most $1+(\delta/8)^2 \le 1+\delta^2/2$ (and at least $1-\delta^2$) and $x$-coordinate at most $\delta/2$ (and at least $-\delta/2$). Thus, their squared distance to $s_Q = (-\delta/2,\delta^2)$ is at most 
  $$ \big(-\delta/2 - \delta/2\big)^2 + \big((1 + \delta^2/2) - \delta^2\big)^2 = \delta^2/16 + (1 - \delta^2 + \delta^4/4) \le 1, $$
  proving \ref{U1}. For \ref{U2} we similarly use that all vertices on $Q$, except for $s_Q^*$ and $t_Q^*$ have $y$-coordinate at least $-(\delta/8)^2$ to obtain that $s_P$ is within distance 1 of these vertices. The distance between $s_P$ and $s_Q^*$ is exactly 1 by definition. The only exception is $t_Q^*$, and indeed the distance between $s_P$ and $t_Q^*$ is larger than 1. Thus, $s_P$ is in distance 1 of any point on $Q$ except for $t_Q^*$. Claim \ref{U3} is symmetric, and \ref{U4} follows directly from the definitions.
  \end{proof}
  
  For the other direction, assume that there is a traversal $\tau$ of $P,Q$ staying in distance 1. We want to show that $\dfr(P_i,Q_j) \le 1$ holds for some $i,j$. Consider any time step where $\tau$ is in $s_Q^*$ on $Q$ and some vertex $x$ on $P$. By claim \ref{L1} below, $x$ has to be a vertex $s_P^{(i)}$, for some $1 \le i \le k$. We may stay in $s_P^{(i)}$ for some steps, until we at some point in time make a step from $s_P^{(i)}$ to $b_P^{(i)}$. We show that at this time we have to be in $b_Q^{(j)}$ for some $j$. Consider the possible vertices of $Q$ where we can be. By \ref{L1} and \ref{L2} the points $s_Q^*$ and $t_Q^*$ are too far from $b_P^{(i)}$. 
  The vertices $s_Q$ and $t_Q$ are both in distance 1 of $b_P^{(i)}$, however, we already visited $s_Q^*$, and since $s_Q$ comes before $s_Q^*$ on $Q$ we cannot be at $s_Q$. Moreover, since the point in time at which we were at $(s_P^{(i)}, s_Q^*)$ we only visited the vertices $s_P^{(i)}$ and $b_P^{(i)}$ on $P$, and both have distance more than~1 to the bottleneck vertex $t_Q^*$, so we cannot walk to $t_Q$, which comes after $t_Q^*$ on $Q$. Hence, while being at $b_Q^{(i)}$ we cannot be at $s_Q$ or $t_Q$. The vertices of $Q_\ell$, for any $\ell$, are also not in distance 1, by \ref{L3}. Finally, by \ref{L5} we cannot be at $e_Q$.
  Hence, while vertex at $b_P^{(i)}$ we have to be at $b_Q^{(j)}$ for some $j$. By \ref{L3} and \ref{L4}, the next step is simultaneous in $P$ and $Q$, since the distance from $b_P$ to $Q_j$ as well as from $b_Q$ to $P_i$ is larger than~1. After this simultaneous step we are at the starting vertices of $P_i$ and $Q_j$. By \ref{L3} and \ref{L4}, we cannot make a step to $e_P$ while staying in $Q_j$ (or to $e_Q$ while staying in $P_i$), since their distance is too large. Hence, $\tau$ contains a traversal $\tau'$ of $P_i,Q_j$, after which it makes a simultaneous step to $e_P^{(i)},e_Q^{(j)}$. Since $\tau$ stays in distance 1, also $\tau'$ stays in distance 1, and we obtain $\dfr(P_i,Q_j) \le 1$ (for some $i,j$). It remains to show the following.
  
  \begin{claim} We have the following lower bounds on vertex distances:
  \begin{enumerate}[label=(L\arabic*)]
    \item $s_Q^*$ has distance larger than 1 to any vertex of $P$ except $s_P$, \label{L1}
    \item $t_Q^*$ has distance larger than 1 to any vertex of $P$ except $t_P$, \label{L2}
    \item $b_P$ and $e_P$ have distance larger than 1 to any vertex of $Q_\ell$ for all $\ell$, \label{L3}
    \item $b_Q$ and $e_Q$ have distance larger than 1 to any vertex of $P_\ell$ for all $\ell$, \label{L4}
    \item $b_P$ and $e_Q$ as well as $e_P$ and $b_Q$ have distance larger than 1, \label{L5}
  \end{enumerate}
  \end{claim}
  \begin{proof}
     For \ref{L1}, since $s_Q^* = (-\delta/2,-\delta^2)$ and all vertices of $P$ have $y$-coordinate at least $1-\delta^2$, vertices in distance 1 to $s_Q^*$ on $P$ are of the form $(-\delta/2,1-\delta^2)$, which is $s_P$. \ref{L2} is symmetric.
     
     For \ref{L3}, since $P_\ell,Q_\ell$ are $\delta/8$-placed, all vertices of $Q_\ell$ have $x$-coordinate at least $-\delta/8$ and $y$-coordinate at most $(\delta/8)^2$. Thus, the squared distance from any point on $Q_\ell$ to $b_P = (-\delta/2,1)$ is
  $$ \ge \big( \delta/2 - \delta/8 \big)^2 + \big( 1 - \delta^2/64 \big)^2   \ge \delta^2/16 + ( 1 - \delta^2/32 + \delta^4 / 4096 )  > 1. $$
  The distance to $e_P$ is symmetric, as is \ref{L4}. Finally, \ref{L5} follows directly from the definitions.
  \end{proof}
\end{proof}

\subsection{Implementing Gates} \label{sec:gates}

It remains to prove \lemref{gates}, i.e. to implement the different types of gates. Let $\delta \in (0,1)$ and fix vectors $a \in A, b \in B$.

\paragraph*{Comparison Gate}
First consider an input gate $g$ to formula $\F$, i.e., a comparison $[a_i \le b_i]$ for some $i$. We define $P = P_\delta(g,a)$ and $Q = Q_\delta(g,b)$ as degenerate curves consisting only of a single vertex. Specifically, we set $P_\delta(g,a) := \big( (0,1+\delta^2 a_i / M) \big)$ and $Q_\delta(g,b) := \big( (0,\delta^2 b_i / M) \big)$. 
\begin{lem}[Comparison Gate] \label{lem:comparison}
  The curves $P_\delta(g,a),Q_\delta(g,b)$ satisfy \ref{P1} and \ref{P2}.
\end{lem}
\begin{proof}
Note that the \Fr distance of degenerate curves consisting of one vertex is simply the Euclidean distance of the vertices. Thus, $\dfr(P,Q) = |1 + \delta^2 a_i / M - \delta^2 b_i / M|$. Since $\delta \in (0,1)$ and $a_i,b_i \in \{-M,\ldots,M\}$, this equals $1 + \delta^2 (a_i-b_i) / M$, which is at most 1 if and only if $a_i \le b_i$. This shows property \ref{P1}, and property \ref{P2} holds by definition.
\end{proof}

\paragraph*{AND Gate}
Now consider an AND gate  $g$ in $\F$, and let $g_1,\ldots,g_k$ be the children of $g$ in $\F$. We (inductively) construct the curves $P_\ell := P_{\delta'}(g_\ell,a)$ and $Q_\ell := Q_{\delta'}(g_\ell,b)$ for $1 \le \ell \le k$ and some $\delta' \le \delta/4$. The only facts about $P_\ell,Q_\ell$ that we will use are property \ref{P1} and that $P_\ell,Q_\ell$ are $\delta/4$-placed (by \ref{P2} and $\delta' \le \delta/4$).
We translate the curves $P_\ell$ and $Q_\ell$ along the $x$-axis by $+\delta/2$, if $\ell$ is odd, and by $-\delta/2$, if $\ell$ is even, i.e. we add $\delta/2$ or $-\delta/2$ to the $x$-coordinate of every vertex of $P_\ell$ and $Q_\ell$. This results in curves $P_\ell', Q_\ell'$. Note that translating both curves does not change the \Fr distance and thus we have $\dfr(P'_\ell,Q'_\ell) \le 1$ if and only if $g_\ell$ evaluates to true, by \ref{P1}. We form the final curve $P = P_\delta(g,a)$ by concatenating $P'_1,\ldots,P'_k$, and similarly we form $Q = Q_\delta(g,b)$ by concatenating $Q'_1,\ldots,Q'_k$. 
\begin{lem}[AND Gate] \label{lem:and}
  The curves $P_\delta(g,a),Q_\delta(g,b)$ satisfy \ref{P1} and \ref{P2}.
\end{lem}
\begin{proof}
The curves $P_\ell,Q_\ell$ are $\delta/4$-placed since $\delta' \le \delta/4$. The translation by $\pm \delta/2$ is sufficiently small to ensure that $P'_\ell,Q'_\ell$ and thus also $P,Q$ are $\delta$-placed, so property \ref{P2} holds.

For property \ref{P1}, first note that if all gates $g_i$ evaluate to true then $\dfr(P'_i,Q'_i) \le 1$ for all $i$, so we can traverse $P'_i$ and $Q'_i$ staying in distance 1, and concatenating these traversals for $1 \le i \le k$ (with a simultaneous step in $P$ and $Q$ between any $P'_i,Q'_i$ and $P'_{i+1},Q'_{i+1}$) yields a traversal of $P,Q$ staying in distance 1. Thus, if $g$ evaluates to true then $\dfr(P,Q) \le 1$. 

For the other direction, we first show that any vertices in $P'_i$ and $Q'_j$ have distance larger than~1 for $i \not\equiv j \pmod 2$. For simplicity, say that $i$ is odd and $j$ is even. Since we translated $P_i$ by $+\delta/2$ and $Q_j$ by $-\delta/2$, we obtain that $P'_i$ is contained in $[\delta/2 - \delta/4,\delta/2+\delta/4] \times [1-(\delta/4)^2,1+(\delta/4)^2]$ and $Q'_j$ is contained in $[-\delta/2 - \delta/4,-\delta/2+\delta/4] \times [-(\delta/4)^2,(\delta/4)^2]$. Thus, a lower bound for the distance between any two vertices of $P'_i$ and $Q'_j$ is the distance from the lower left corner of the region containing $P'_i$ and the upper right corner of the region containing $Q'_j$, which is 
$$\left(\big((\delta/2 - \delta/4) - (-\delta/2 + \delta/4)\big)^2 + \big(1-2(\delta/4)^2\big)^2  \right)^{1/2} = \left(\delta^2/4 + 1 - 4(\delta/4)^2 + 4(\delta/4)^4\right)^{1/2} > 1.$$ 
Hence, no traversal staying in distance 1 can be simultaneously in $P'_i$ and $Q'_j$, for $i \not\equiv j \pmod 2$.

Now assume that $\dfr(P,Q) \le 1$ and consider a traversal $\tau$ realizing the \Fr distance.  
Note that $\tau$ starts in the first vertices of $P'_1,Q'_1$, and consider the first time where the traversal reaches a vertex in $P'_2$ or $Q'_2$. Observe that the previous move has to be a simultaneous move from the last vertices in $P'_1,Q'_1$ to the first vertices in $P'_2,Q'_2$, since we cannot be in $P'_1$ and $Q'_2$ at the same time, or symmetrically in $P'_2$ and $Q'_1$, as $1 \not\equiv 2 \pmod 2$. Thus, we can split the traversal $\tau$ into a traversal of $P'_1,Q'_1$ and a traversal of the remainder of $P,Q$. Proceeding inductively, we can split $\tau$ into traversals $\tau_i$ of $P'_i,Q'_i$ for $1 \le i \le k$. Since $\tau$ stays in distance 1, each $\tau_i$ also stays in distance~1. This shows $\dfr(P'_i,Q'_i) \le 1$, and by $\dfr(P'_i,Q'_i) = \dfr(P_i,Q_i)$ and property \ref{P1} we obtain that all gates $g_i$ evaluate to true. This proves correctness of the AND gate construction.
\end{proof}

\paragraph*{OR Gate}
We simulate OR gates by a combination of all previous constructions in \secref{frechet}. 
Consider any OR gate $g$ in $\F$, and let $g_1,\ldots,g_k$ be its children in $\F$, corresponding to subformulas $\F_{1},\ldots,\F_{k}$. 
Note that the outer OR construction from \lemref{outeror} does not suffice to simulate an OR gate: If we recursively build curves $P_\ell,Q_\ell$ simulating subformula $\F_\ell$ and combine them with an outer OR then it is possible to pair up $P_i$ and $Q_j$ for $i \ne j$, thus not respecting the structure of the formula. To prevent these mismatches, we adapt the subformulas $\F_\ell$ as follows.
We first present a sketch of this adaptation, for the details see \lemref{adaptation}.
We consider the auxiliary formulas $\F'_{\ell} = [\ell \le \ell]  \wedge \F_{\ell} \wedge [-\ell \le -\ell]$, and construct the curves $P'_\ell, Q'_\ell$ that our conversion yields when run on $\F'_\ell$. 
This might seem like an unnecesasry step at first sight, since the comparisons $[\ell \le \ell], [-\ell \le -\ell]$ are always true, so $\F'_\ell$ simplifies to $\F_\ell$. However, consider $i \ne j$ and the constructed curves $P'_i$ and $Q'_j$. Then the comparison $[\ell \le \ell]$ becomes $[i \le j]$, since the left operand of the comparison is encoded in $P$ and the right operand is encoded in $Q$.  
Similarly, the comparsion $[-\ell \le -\ell]$ becomes $[-i \le -j]$. One of these comparisons is wrong for $i \ne j$, so that the AND evaluates to false. Thus, the \Fr distance of $P'_i, Q'_j$ is larger than 1 for any $i \ne j$. 

This trick allows us to use the outer OR construction to finish our OR gate construction. Indeed, running \lemref{outeror} on the curves $P'_1,Q'_1,\ldots,P'_k,Q'_k$ yields curves $P,Q$ such that $\dfr(P,Q) \le 1$ if and only if there are $i,j$ with $\dfr(P'_i,Q'_j) \le 1$. Since by \lemref{adaptation}.(1) we must have $i = j$ for \Fr distance at most 1, we have $\dfr(P,Q) \le 1$ if and only if $\dfr(P'_\ell,Q'_\ell) \le 1$ for some $\ell$. Finally, by \lemref{adaptation}.(2) we have $\dfr(P,Q) \le 1$ if and only if $\F_\ell$ evaluates to true for some $\ell$. Hence, we correctly simulated the OR gate $g$.

\begin{lem} \label{lem:adaptation}
  We can construct $\delta/8$-placed curves $P'_1,Q'_1,\ldots,P'_k,Q'_k$ satisfying (1) $\dfr(P'_i,Q'_j) > 1$ for $i \ne j$ and (2) $\dfr(P'_\ell,Q'_\ell) \le 1$ if and only if subformula $\F_\ell$ evaluates to true on $(a,b)$.
\end{lem}
\begin{proof}
As above, $P'_\ell,Q'_\ell$ are the curves produced by our conversion when run on $\F'_\ell = [\ell \le \ell] \wedge \F_\ell \wedge [-\ell \le -\ell]$. More precisely, we recursively construct the curves $P_\ell := P_{\delta/32}(g_\ell,a)$ and $Q_\ell := Q_{\delta/32}(g_\ell,b)$ for $1 \le \ell \le k$. For the comparison $[\ell \le \ell]$ we construct the degenerate curves $C^P_\ell = \big( (0,1+\delta/32 \cdot \ell / n) \big)$ and $C^Q_\ell = \big( (0,\delta/32 \cdot \ell / n) \big)$ (see \lemref{comparison}), and for the comparison $[-\ell \le -\ell]$ we construct the degenerate curves $D^P_\ell = \big( (0,1-\delta/32 \cdot \ell / n) \big)$ and $D^Q_\ell = \big( (0,-\delta/32 \cdot \ell / n) \big)$. Note that $C^P_\ell,C^Q_\ell, D^P_\ell,D^Q_\ell, P_\ell,Q_\ell$ are $\delta/32$-placed, so we can apply the AND gate construction resulting in $\delta/8$-placed curves, see \lemref{and}. 
Specifically, we translate $C^P_\ell, C^Q_\ell, D^P_\ell, D^Q_\ell$ by $+\delta/16$ along the $x$-axis, and $P_\ell,Q_\ell$ by $-\delta/16$. Finally, we concatenate (the translated versions of) $C^P_\ell, P_\ell, D^P_\ell$ to obtain $P'_\ell$ and we concatenate (the translated versions of) $C^Q_\ell, Q_\ell, D^Q_\ell$ to obtain $Q'_\ell$. 
Note that $P'_\ell,Q'_\ell$ are $\delta/8$-placed, so we can indeed use \lemref{outeror} to combine $P'_1,Q'_1,\ldots,P'_k,Q'_k$ to $\delta$-placed curves $P,Q$ as in the paragraph before the lemma. 

From the correctness of the AND gate and comparison constructions we obtain that $P'_\ell,Q'_\ell$ simulate $\F'_\ell$, i.e. $\dfr(P'_\ell,Q'_\ell) \le 1$ if and only if $\F'_\ell(a,b)$ evaluates to true. Since $\F'_\ell$ simplifies to $\F_\ell$, this yields (2).
For (1), consider $P'_i,Q'_j$ for $i \ne j$. One of the comparisons $[i \le j], [-i \le -j]$ is wrong, so we obtain $\dfr(C^P_i,C^Q_j) > 1$ or $\dfr(D^P_i,D^Q_j) > 1$. Since $C^P_i,C^Q_j$ and $D^P_i,D^Q_j$ and $P_i,Q_j$ are $\delta/32$-placed (before we translate them), the correctness argument of the AND gate, \lemref{and}, yields $\dfr(P'_i,Q'_j) > 1$, as desired. 
\end{proof}

\paragraph*{Finishing the Proof}
We obtain \lemref{gates} by putting together the gate constructions of this section. This yields a recursive construction of curves $P_\delta(g,a), Q_\delta(g,a)$ simulating gate $g$, correctness follows from a simple inductive argument. Note that if $g$ has fanin $k$ then we add $\Oh(k)$ vertices to the recursively constructed curves corresponding to the $k$ children, yielding a bound of $\Oh(|\F|)$ on the length of the constructed curves. Finally, for the factor by which we decrease $\delta$ from one gate to its children we note that $1/32$ is sufficiently small. Thus, we may produce $\delta / 32^d$-placed curves for each gate in depth $d$ below $g$. The coordinates of our AND and OR gate constructions thus use $\Oh(\log(1/\delta) + d)$ bits. The only exception is the comparison gate construction, for which we need $\Oh(\log M) = \Oh(\log n)$ additional bits to represent numbers in $\{-M,\ldots,M\}$. Together this yields \lemref{gates}.

\end{document}